\documentclass[1p]
{elsarticle}

\usepackage[usenames,dvipsnames]{xcolor}
\DeclareGraphicsExtensions{.pdf,.jpeg,.png}
\usepackage{epstopdf}
\usepackage{amsmath,amsfonts,amssymb,amsthm}
 \usepackage[tight,footnotesize]{subfigure}
 \usepackage{bbm}
\usepackage{bm}

\usepackage{mathrsfs}
\usepackage{dsfont}
\usepackage{amsbsy}
\usepackage{stmaryrd}
\usepackage{url,algorithm,algpseudocode}
\usepackage{tikz}
\usetikzlibrary{arrows,positioning,automata, backgrounds, decorations.markings}

\usepackage{cancel}

\newcommand{\bbR}{\mathbb{R}}
\newcommand{\bbP}{\mathbb{P}}
\newcommand{\bbE}{\mathbb{E}}

\newcommand{\eps}{\varepsilon}

\newcommand{\vr}[1]{\mathbf{#1}}

\def\qed{$\hfill\blacksquare$\newline}

\newcommand{\calC}{\mathcal{C}}

\newcommand{\calB}{\mathcal{B}}

\newcommand{\calV}{\mathcal{V}}

\newcommand{\X}{\mathbf{X}}

\newcommand{\x}{\mathbf{x}}

\newcommand{\tY}{\tilde{Y}}
\newcommand{\hY}{\hat{Y}}

\newcommand{\size}{N}
\newcommand{\upsize}{^{(N)}}
\newcommand{\N}{^{(N)}}

\newcommand{\class}{\mathscr{A}} 
\newcommand{\cstsp}{S} 
\newcommand{\ctrsp}{E} 
\newcommand{\lbset}{\mathscr{L}} 

\newcommand{\loctr}[1]{\epsilon_{#1}} 
\newcommand{\loclab}[1]{\alpha_{#1}} 
\newcommand{\locarr}[1]{s_{#1} \xrightarrow{\loclab{i}} s'_{#1}} 

\newcommand{\astsp}{Y} 

\newcommand{\pop}{\mathcal{X}} 
\newcommand{\popn}{\hat{\mathcal{X}}} 
\newcommand{\pstsp}{\mathcal{S}} 
\newcommand{\ptrsp}{\mathcal{T}} 
\newcommand{\pntrsp}{\hat{\mathcal{T}}} 
\newcommand{\pstvec}{\mathbf{X}} 
\newcommand{\pnstvec}{\hat{\mathbf{X}}} 
\newcommand{\pinst}{\mathbf{x}_0} 
\newcommand{\pninst}{\hat{\mathbf{x}}_0} 
\newcommand{\syncset}[1]{\mathbb{S}_{#1}}

\newcommand{\drift}{\textbf{F}} 
\newcommand{\fluid}{\boldsymbol\Phi} 
\newcommand{\diff}{\textbf{G}} 
\newcommand{\linear}{\textbf{Z}}
\newcommand{\mean}{\mathbf{E}}
\newcommand{\cov}{\mathbf{C}}
\newcommand{\jac}{\mathbf{J}}

\newcommand{\dta}{\mathbb{D}}
\newcommand{\gdta}{\dta}
\newcommand{\alp}{\Sigma}
\newcommand{\atprop}{\Gamma_{\pstsp}} 
\newcommand{\gstsp}{Q}
\newcommand{\ginst}{q_0}
\newcommand{\gfinst}{F}
\newcommand{\gtr}{\rightarrow}

\newcommand{\tr}[2]{\xrightarrow{#2}_{#1}}

\newcommand{\aca}[1]{\mathscr{P}_{I_{#1}}}
\newcommand{\acstsp}{\hat{\cstsp}}
\newcommand{\actrsp}{\hat{\ctrsp}}
\newcommand{\gprop}[3]{#1 \in [#2,#3]}
\newcommand{\gpropg}[2]{#1 \geq  #2 }
\newcommand{\gpropl}[2]{#1 \leq #2}
\newcommand{\gp}[3]{\mathcal{P}_{ #3}^{#2}(#1) }
\newcommand{\gpop}{\boldsymbol\pop}
\newcommand{\apro}{{\mathscr{P}}}

\renewcommand{\P}{\mathbf{P}}

\theoremstyle{plain}
\newtheorem{theorem}{Theorem}[section]

\newtheorem{proposition}{Proposition}[section]

\theoremstyle{definition}
\newtheorem{definition}{Definition}[section]
\newtheorem{example}{Example}[section]
\theoremstyle{remark}

\newtheorem{remark}{Remark}[section]

\begin{document}


\journal{Information and Computation}

\begin{frontmatter}

\title{Model Checking Markov Population Models by Stochastic Approximations}

\author[TS,SA]{Luca Bortolussi}
\address[TS]{DMG, University of Trieste, Trieste, Italy.}
\address[SA]{MOSI, Department of Computer Science, Saarland University, Saarbr\"ucken, Germany.}
\ead{luca@dmi.units.it}
\author[LU]{Roberta Lanciani}
\author[LU]{Laura Nenzi}
\address[LU]{IMT, Lucca, Italy.}
\ead{roberta.lanciani@imtlucca.it}
\ead{laura.nenzi@imtlucca.it}

\def\titlerunning{Checking MPM by Stochastic Approximation}
\def\authorrunning{L. Bortolussi \& R. Lanciani \& L. Nenzi}

\maketitle


\begin{abstract}
Many complex systems can be described by population models, in which a pool of agents interacts and produces complex collective behaviours.   We consider the problem of verifying formal properties of the underlying mathematical representation of these models, which is a Continuous Time Markov Chain, often with a huge state space. 
To circumvent the state space explosion, we rely on stochastic approximation techniques, which replace the large model  by a simpler one, guaranteed to be probabilistically consistent. We show how to  efficiently and accurately verify properties of random individual agents, specified by  Continuous Stochastic Logic extended with Timed Automata (CSL-TA), and how to lift these specifications to the collective level, approximating the number of agents satisfying them using second or higher order stochastic approximation techniques.  
\end{abstract}


\end{frontmatter}




\section{Introduction}
\label{sec:intro}
Many real-life examples of large complex systems, ranging from (natural) biochemical pathways to (artificial) computer networks, exhibit \textit{collective behaviours}. These global dynamics are the result of intricate interactions between the large number of individual entities that comprise the populations of these systems. Understanding, predicting and controlling these emergent behaviours is becoming an increasingly important challenge for the scientists of the modern era. In particular, the development of an efficient and well-founded mathematical and computational modelling framework is essential to master the analysis of such complex collective systems.

In the Formal Methods community, powerful automatic verification techniques have been developed to validate the performance of a model of a system. In such \textit{model checkers} \cite{mcbook}, the model and a property of interest are given in input to an algorithm which verifies whether or not the requirement is satisfied by the representation of the system. 
As the dynamics of a collective system  is intrinsically subject to noisy behaviours, especially when the population is not very large, the formal analysis and verification of a collective system have to rely on appropriate \textit{Stochastic Model Checking} techniques. For instance, in \cite{ctmcmc}, Continuous Stochastic Logic formulae are checked against models of the system expressed as Continuous Time Markov Chains (CTMC, \cite{norris}), which are a natural mathematical framework for population models. These approaches  are based on an
exhaustive exploration of the state space of the model, which limits their practical use, due to  \textit{state space explosion}: when the number of interacting agents in the population increases, the number of states of the underlying CTMC quickly reaches astronomical values. 
 To deal with this problem, some of the most successful applications of Stochastic Model Checking to large population models are based on statistical analysis \cite{prism,jha2009statistical,bortolussi2016smoothed}, which still remain costly from a computational point of view, because of the need or running simulation algorithms a large number of times.

In this work, we take a different approach, exploiting a powerful class of methods to accurately approximate the dynamics of the individuals and the population, that goes under the name of \emph{Stochastic Approximations} \cite{tutorial}. 
%
%

\vspace{3mm}
\noindent\textbf{Related Work.} 
Stochastic approximation methods have been successfully used in the computational biochemistry community \cite{grima2010, vankampen} to approximate the noisy behaviour of collective systems by a simpler process whose behaviour can be extracted by solving a (numerically integrable) set of \textit{Differential Equations} (DEs), resulting in a fast and easy way of obtaining an estimation of the dynamics of the model. 
Moreover, for almost all the techniques that we are going to consider in this work, the quality of the estimations improves as the number of agents in the system increases, keeping constant the computational cost and reaching exactness in the limit of an infinite population. In this way, such approximation methods actually take advantage of the large sizes of the collective systems, making them a fast, accurate and reliable approach to deal with the curse of the state space explosion. Among the many types of Stochastic Approximations present in the literature, we are going to exploit the \textit{Fluid Approximation} (FA) \cite{tutorial,fluidmc}, the \textit{Central Limit Approximation} (CLA) \cite{vankampen, kurtz}, and the  \textit{System Size Expansion} (SSE) \cite{schnoerr2016approximation}. We are also  going to use \textit{Moment Closure} (MC) \cite{schnoerr2016approximation} combined with distribution reconstruction techniques based on the \textit{Maximum Entropy} principle \cite{andreychenko2015model}.

Stochastic Approximations entered into the model checking scene only recently. Pioneering work focussed on checking CSL properties \cite{concur12,fluidmc,bertinoro13} or deterministic automata specifications \cite{HaydenTCS, HaydenTSE} for a single random individual in a population. Following this line of  work, more complex individual properties  had been considered, in particular rewards \cite{qapl15} and timed automata with one clock \cite{formats15}. 
Another direction of integration of stochastic approximations and model checking is related to the so called local-to-global specifications \cite{qest}, in which individual properties, specified by  timed automata (with some restrictions), are lifted at the collective level by counting how many agents satisfy a local specification. This lifting is obtained by applying the CLA to approximate the distribution of agents \cite{qest}   or by moment closure to obtain bounds \cite{HaydenTCS, HaydenTSE}. A simpler approach, focussing on expected values at the collective level, is \cite{anjaL2G}.  
Finally, stochastic approximation has been used also to approximate global reachability properties, either exploiting central limit results for hitting times \cite{epew}, or by a clever discretisation of the Gaussian processes obtained by the CLA \cite{qest16}. 

\vspace{3mm}
\noindent\textbf{Contributions.} 
In this paper, we start from the approach of \cite{qest} for the approximation of satisfaction probabilities of local-to-global properties, and extend it in several directions:
\begin{itemize}
\item We extend fluid model checking \cite{fluidmc} to a subset of CSL-TA \cite{cslta}, a logic specifying temporal properties by means of Deterministic Timed Automata (DTA) with a single clock. We consider  in particular DTAs in which the clock is never reset, and provide a model checking algorithm also for nested formulae, leveraging fluid approximation.
\item We lift CSL-TA properties to the collective level, exploiting the central limit approximation, thus extending the approach of \cite{qest} to a more complex set of properties. We also remove some restrictions on the class of models considered with respect to those discussed in \cite{qest}. 
\item We extend both \cite{qest}  and \cite{fluidmc} by showing how to effectively use higher order approximations to correct for finite size effects. This requires to integrate within the model checking framework either moment closure or higher-order SSE \cite{schnoerr2016approximation}, together with maximum entropy distribution reconstruction \cite{andreychenko2015model}.
\end{itemize}
Throughout the paper, we make use of a simple but instructive running example of an epidemic model to illustrate the presented techniques.


\vspace{3mm}
\noindent\textbf{Paper Structure.}  
The paper is organised as follows. In Section \ref{sec:popmod}, we introduce the class of models we consider, and in Section \ref{sec:property}, the property specification language. Section \ref{sec:cla} contains an introduction on  stochastic approximation techniques.  Section  \ref{sec:checkCSLTA} shows how to model check local properties described by CSL-TA, while Section \ref{sec:modelchecking} deals with local-to-global properties. Conclusions are drawn in Section \ref{sec:conc}. The appendix contains novel proofs. 

%

\section{Markov Population Models}
\label{sec:popmod}

In this section, we introduce a formalism to specify \emph{Markovian Population Models}. These models  consists of typically large collections of interacting components, or \textit{agents}. 
Each component  is a finite state machine, which can change internal state by interacting  with other agents or with the environment. Agents can be of different kinds or classes. Interactions are described by specifying which kinds of agents participate in the interaction and the rate at which it happens. The rate is a function of the collective state of the model, i.e. of the population size. This information, together with an initial state, describe the full population model and it defines a Markov chain in continuous time. 

More specifically, an agent class  $\class$  defines its (finite) state space and its (finite) set of \textit{local} transitions. In the following, the descriptor '\textit{local}' refers to the fact we are formalizing the model at the agent level, whereas we use '\textit{global}' to define state spaces and transitions when modelling the entire population.
\begin{definition}[Agent class]
\label{class}
An \emph{agent class} $\class$ is a pair $(\cstsp, \ctrsp)$ where $\cstsp = \{s_1, \ldots, s_n\}$ is the state space of the agent and $\ctrsp = \{\loctr{1}, \ldots, \loctr{m}\}$ is the set of local transitions of the form $\loctr{i} = \locarr{i}$, $i \in \{1, \ldots, m\}$, where $\loclab{i}$ is the transition label, taken from the label set $\lbset$. 
\end{definition}
The label $\loclab{i}$ of a local transition $\loctr{i} = \locarr{i}$ may not be unique. Without loss of generality, we require that  two local transitions having the same initial and final states must have different labels (if not, just rename labels). 
An agent belonging to the class $\class_h = (\cstsp_h, \ctrsp_h)$ is identified by a random variable $\astsp_h(t) \in \cstsp$, denoting the state of the agent at time $t$, and the initial state $\astsp_h(0) \in \cstsp$.
%

\begin{figure}[!t]
\begin{center}
\begin{tikzpicture}[on grid, shorten <=1pt, >=stealth', auto]
  \node at (2.5,0)   [state] (R)   {{\bf R}};
  \node at (0,0)   [state] (I)   {{\bf I}};
  \node at (1.25,2) [state] (S)   {{\bf S}};
  
  \path [->] (I)  edge [loop left, left]   node {$inf$}    ()
                        edge [bend right, above]  node {$patch_1$}    (R)
                 (R)  edge [bend right, right]       node {$loss$}   (S)
                 (S)  edge [bend right, left]     node {$inf$}    (I)
                        edge [right]                         node {$ext$}    (I)
                       edge [left, near end]      node {$patch_0$} (R);
\end{tikzpicture}
\end{center}
\caption{ The automaton representation of a network node.}
\label{SIRagent}
\end{figure}
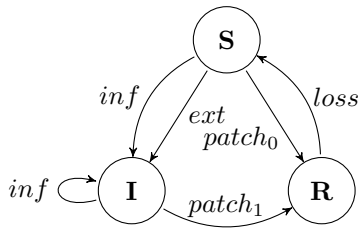

\medskip

\noindent\textbf{Running Example.} 
In order to illustrate the definitions and the techniques of the paper, we consider a simple example of  a worm  epidemic in  a peer-to-peer network composed of $N$ nodes (see e.g.\ \cite{meanfieldP2P} for mean field analysis of network epidemics). Each node is modelled by the simple agent shown in Figure \ref{SIRagent}, which has three states: susceptible to infection (S), infected (I), and patched/immune to infection (R). 
The contagion of a susceptible node can occur due to an event external to the network ($ext$), like the reception of an infected email, or by file sharing with an infected node within the network ($inf$). Nodes can also be patched, at different rates, depending if they are infected ($patch_1$) or not ($patch_0$). A patched node remains immune from the worm for some time, until immunity is lost ($loss$), modelling for instance the appearance of a new version of the worm. 

The agent class $\class_{node} = (\cstsp_{node}, \ctrsp_{node})$ of the network node can be easily reconstructed form the automaton representation in Figure \ref{SIRagent}:  its local states are $\cstsp_{node} = \{s_S,s_I,s_R\}$, which we also denote as $\{S,I,R\}$, and its local transitions are $\ctrsp_{node} =\{S\xrightarrow{inf} I, S\xrightarrow{ext} I, I\xrightarrow{inf} I, I\xrightarrow{patch_1} R, S\xrightarrow{patch_0} R, R\xrightarrow{loss} S \}$.

\medskip

In the following, without loss of generality, we consider populations of $N$ agents $\astsp\upsize_k$, $k \in \{1, \ldots, N\}$, all belonging to the same class $\class = (\cstsp, \ctrsp)$ with $S = \{s_1, \ldots, s_n\}$.  All the results we will present in the following hold for models with multiple classes of agents, at the price of keeping an extra index to identify the class of each agent.  
We further make the classical assumption that agents in the same state are indistinguishable, hence the state of the population model can be described by \textit{collective} or  \textit{counting variables} $\pstvec\upsize = (X\upsize_{1}, \ldots, X\upsize_{n})$, $X\upsize_j \in \{0, \ldots, N\}$, defined by $X\upsize_{j} = \sum^{N}_{k = 1} \mathds{1}\{Y\upsize_{k} = j\}$. 
The initial state $\x_0\upsize$ is given by $\x_0\upsize = \X\upsize(0)$, and the counting variables satisfy the conservation relation $\sum_{j \in \cstsp} X\upsize_{j} = N$. To complete the definition of a population model, we need to specify its \emph{global} transitions, describing all the possible events that can change the state of the system.

\begin{definition}[Population model] 
\label{populationModel}
A \textit{population model} $\pop\upsize$ of size $N$ is a tuple $\pop\upsize = (\class,  \ptrsp\upsize, \pinst\upsize)$, where:
\begin{itemize}
\item $\class$ is an agent class, as in Definition \ref{class};
\item $\ptrsp\upsize = \{ \tau_1, \ldots, \tau_\ell \}$ is the set of global transitions of the form $\tau_i = (\syncset{i}, f\upsize_i)$, where:
\begin{itemize}
\item $\syncset{i} = \{ s_1 \xrightarrow{\loclab{1}} s'_1, \ldots, s_p \xrightarrow{\loclab{p}} s'_p \}$ is the (finite) set of local transitions synchronized by $\tau_i$;
\item $f\upsize_{i}: \mathds{R}^n \longrightarrow \mathds{R}_{\geq 0}$ is the  (Lipschitz continuous) global rate function. 
\end{itemize}
\item$\pinst\upsize$ is the initial state.
\end{itemize}
\end{definition}
The rate $f\upsize_i$ gives the expected frequency of transition $\tau_i$ as a function of the state of the system. We assume $f\upsize_i$ equal to zero if there are not enough agents available to perform the transition. The synchronization set $\syncset{i}$, instead, specifies how many agents are involved in the transition $\tau_i$ and how they change state: when $\tau_i$ occurs, we see the local transitions $s_1 \xrightarrow{\loclab{1}} s'_1, \ldots, s_p \xrightarrow{\loclab{p}} s'_p$ fire at the (local) level of the $p$ agents involved in $\tau_i$. 

\begin{figure}[!t]
\begin{center}
\begin{tabular}{|ccccc|}
\hline
$\tau_{ext}$ & \phantom{aaaa} & $\{S \xrightarrow{ext} I\}$ & \phantom{aaaa} & $f_{ext}(\X ) = \kappa_{ext} X_S$ \\
$\tau_{loss}$ & & $\{R \xrightarrow{loss} S\}$ & & $f_{loss}(\X ) = \kappa_{loss} X_R$\\ 
$\tau_{patch_0}$  & & $\{S \xrightarrow{patch_0} R\}$ & & $f_{patch_0}(\X ) = \kappa_{patch_0} X_S$\\ 
$\tau_{patch_1}$ & & $\{I \xrightarrow{patch_1} R\}$ & & $f_{patch_1}(\X ) = \kappa_{patch_1} X_I$\\ 
$\tau_{inf}$ & & $\{S \xrightarrow{inf} I, I \xrightarrow{inf} I\}$ & & $f_{inf}(\X ) =  \frac{1}{N} \kappa_{inf} X_S X_I$\\ 
\hline
\end{tabular}
\end{center}
\caption{Global transitions of the network epidemic model}
\label{fig:SIRtransitions}
\end{figure}

\medskip

\begin{example} \label{ex:popmodel}
The population model $\pop\upsize_{net} = (\class_{node},  \ptrsp\upsize, \pinst\upsize)$ for the epidemic example has population variables $\X = (X_S, X_I, X_R)$. The initial conditions $\pinst\upsize$ are simply a network of susceptible nodes,  $\pinst\upsize = (N,0,0)$.  The set $\ptrsp\upsize$, instead, specifies five global transitions: $\tau_{ext}, \tau_{loss}, \tau_{patch_0}, \tau_{patch_1}, \tau_{inf} \in \ptrsp\upsize$, detailed in Figure~\ref{fig:SIRtransitions}. \\
As an example, consider the transition $\tau_{ext}$ encoding the external infection. Its synchronisation set $\{S \xrightarrow{ext} I\}$  specifies that only one susceptible node is involved and changes state from $S$ to $I$ at a rate given by $f\upsize_{ext}(\X ) = \kappa_{ext} X_S$, corresponding to a rate of infection $\kappa_{ext}$ per node. 
The transitions $\tau_{loss}, \tau_{patch_0}, \tau_{patch_1}$ have a similar format, while the internal infection i $\tau_{inf}$  involves one $S$-node and one $I$-node, having synchronization set $\{I \xrightarrow{inf} I, S \xrightarrow{inf} I\}$. Furthermore, we assume that an infected node sends infectious messages at rate $\kappa_{inf}$ to a random node, giving a classical density dependent rate function $f\upsize_{inf}(\X ) =  \frac{1}{N} \kappa_{inf} X_S X_I$ \cite{epidbook}.
\end{example}

\medskip

\begin{remark}
\label{rem:restrictions}
The population models of Definition \ref{populationModel} have a main restriction:  the population size is constant. 
This limitation can be removed, as the approximation techniques we will exploit do not rely on it. However, extra care has to be put in treating local properties, as discussed in \cite{fluidmc}.
\end{remark}

Given a population model $\pop\upsize = (\class,  \ptrsp\upsize, \pinst\upsize)$ and a global transition $\tau = (\syncset{\tau}, f\upsize_{\tau}) \in \ptrsp\upsize$ with $\syncset{\tau} = \{ s_1 \xrightarrow{\loclab{1}} s'_1, \ldots, s_p \xrightarrow{\loclab{p}} s'_p \}$, we encode the net change in $\pstvec\upsize$ due to $\tau$ in the {\it update vector} $\textbf{v}_\tau= \sum_{i = 1}^{p} (\textbf{e}_{s_i} - \textbf{e}_{s'_i})$,
%
%
where 
$\textbf{e}_{s_i}$ is the vector that is equal to $1$ in position $s_i$ and zero elsewhere. 

The CTMC $\X\upsize(t)$ associated with $\pop\upsize$ has state space $\pstsp\upsize = \{ (z_1, \ldots, z_{n})\\ \in \mathbb{N}^{n}\ \lvert\ \sum^{n}_{i = 1} z_i = N\}$, initial probability distribution concentrated on $\pinst\upsize$, and \textit{infinitesimal generator matrix} $\textbf{Q}$ defined for $\vr{x},\vr{x'}\in\pstsp\upsize$, $\vr{x} \neq \vr{x'}$, by
\[
q_{\vr{x}, \vr{x'}} = \sum_{\tau \in \ptrsp | \textbf{v}_\tau = \vr{x'} - \vr{x}} f\upsize_\tau(\vr{x}).
\]

Equipped with this definition, we can  analyse a model either by numerical integration of the Kolmogorov equations of the CTMC \cite{norris}, or relying on stochastic simulation and statistical analysis of the sampled trajectories \cite{prism,jha2009statistical}. The first approach is not feasible for large populations ($N$ large), due to the severe state space explosion of this class of models. The second approach scales better, but is still computationally intensive, requiring  many simulations, whose cost typically scales (linearly) with $N$.

\section{Individual and Collective Properties}
\label{sec:property}
We introduce now the class of properties considered in the paper. We distinguish two levels of properties: local properties, describing  the behaviour of individual agents; and global properties, representing the collective behaviour of 
agents with respect to a local property of interest. 
In this respect, our approach is similar to  \cite{anjaL2G,HaydenTCS}.  

Local properties are expressed in terms of a temporal logic. To improve expressiveness of the specifications, we go beyond CSL, as used in \cite{fluidmc,anjaL2G}, and consider CSL-TA \cite{cslta}, an extension of CSL in which path properties are specified by Deterministic Timed Automata (DTA) with a single clock. To rely on fluid approximation techniques, we consider here DTA in which the clock refers always to the global time, i.e it cannot be reset. DTA are used to specify time bounded properties, and we consistently restrict to time-bounded quantification in the CSL layer. This is justified because dealing with steady state properties is always problematic in the context of fluid approximation, see \cite{fluidmc,tutorial,HaydenTCS} for further discussion on this point. 

The global property layer allows us to  specify queries about the fraction of agents that satisfy a given local specification. In particular, given a path or a state formula $\phi$, we want to compute the probability that the fraction  of agents satisfying $\phi$ is smaller or larger than a threshold $\alpha$. This is captured by appropriate operators, that can be then combined  to specify more complex global queries, as in \cite{anjaL2G}. 

Let us fix a population model composed of agents from a class $\class = (\cstsp, \ctrsp)$.  Path properties are specified by a \textit{1-global-clock Deterministic Timed Automata} (1gDTA). The edges of the 1gDTA are labelled by a triple composed of: an action name, taken from the set $\lbset$ of  transition names of the population model;  a boolean formula, interpreted on the states $\cstsp$ of agent $\class$; and a clock constraint, specifying when the transition can fire. The use of actions and formulae to label edges is similar to asCSL \cite{ascsl}, and deviates from the original definition of CSL-TA \cite{cslta}. The intended meaning is that a transition $s \xrightarrow{\alpha} s'$ matches an edge with label $\alpha,\phi$ in the 1gDTA if and only if the action name $\alpha$ is the same and the state $s$ satisfies the formula $\phi$. This allows the specification of more complex path properties 
and provides a mechanism to nest CSL-TA specifications.  
Let us call $\atprop$  this set  of \textit{state propositions on} $\cstsp$, and call $\calB(\atprop)$ the set of boolean combinations over $\atprop$, denoting with $\models_{\atprop}$ the satisfaction relation over $\calB(\atprop)$ formulae, defined in the standard way. We use the letter $\phi$ to range over formulae in $\calB(\atprop)$.
  
%

We consider a single clock variable $x \in \mathds{R}_{\geq 0}$, called \textit{global clock}, never reset in time. Let $\calV$ be the set of \textit{valuations}, i.e. functions $\eta: \{x\} \longrightarrow \mathds{R}^{\geq 0}$ that assign a nonnegative real-value to the global clock $x$, and let  $\calC\calC$ the set of \textit{clock constraints}, which are boolean combinations of basic clock constraints of the form $x \leq a$, where $a \in \mathds{Q}^{\geq 0}$. Finally, we write $\eta(x) \models_{\calC\calC} c$ if and only if $c \in \calC\calC$ is satisfied when its clock variable take the value $\eta(x)$. We are now ready to define 1gDTA.

\begin{definition}[1-global-clock Deterministic Timed Automaton]
\label{def:1gDTA}
A 
1gDTA is specified by  the tuple $\dta = (\lbset, \atprop, \gstsp, \ginst, \gfinst, \gtr)$ where $\lbset$ is the label set of $\class$; $\atprop$ is the set of atomic state propositions; $\gstsp$ is the (finite) set of states of the DTA, with initial state $\ginst \in \gstsp$; $\gfinst \subseteq \gstsp$ is the set of final (or accepting) states, and $\gtr \subseteq \gstsp \times \alp \times \calB(\atprop) \times  \calC\calC \times  \gstsp$ is the edge relation, where $(q, \alpha, \gamma, c, q') \in \gtr$ is usually denoted by $q \xrightarrow{\alpha, \phi, c}q'$. The edges of $\dta$ further satisfy:
\begin{enumerate}
\item (\emph{determinism}) for each $\alpha\in\lbset$,  $s\in\cstsp$ and clock valuation $\eta(x)\in\mathbb{R}_{\geq 0}$, there is exactly one edge $q \xrightarrow{\alpha, \phi, c}q'$ such that $s\models_{\atprop}\phi$ and $\eta(x) \models_{\calC\calC} c$. 
\item (\emph{absorption} of final states) all final states $q\in \gfinst$ are absorbing, i.e. all transitions starting from a final state are self-loops.  
%
%
\end{enumerate}
\end{definition}
When we write a 1gDTA, we do not want to specify all possible transitions in the automaton. Hence, we assume that all non-specified edges are self-loops on the automata state. 
Specifically, given $\alpha$, $s$, and $\eta(x)$, if there is no specified  edge from state $q$ with label $\alpha$, with formula satisfied by $s$ and clock constraint satisfied by $\eta(x)$, then we assume the existence of an edge looping on $q$ and satisfying all  conditions.
The condition of determinism of Definition \ref{def:1gDTA} can be easily enforced by considering 1gDTA that have additional restrictions on the edges, using only formulae $\phi_s$, $s\in\cstsp$, which are true only in state $s$, and requiring that two transitions with label $\phi_s, \alpha$ have mutually exclusive clock constraints. We call these 1gDTA \emph{explicitly deterministic}. All examples of properties in this paper will be of this kind, and it is easy to see that any 1gDTA can be converted into an equivalent explicitly deterministic one, by properly multiplying edges and splitting state formulae and constraints. 



\begin{figure}[!t]
\begin{footnotesize}
\begin{center}
\subfigure[]{
\label{step0}
\begin{tikzpicture}[on grid, shorten <=1pt, >=stealth, auto]
  \node at (2,0)   [initial above,state] (q0)   {$q_0$};
  \node at (0,0)   [state] (qb)   {$q_b$};
  \node at (4,0)   [state, accepting] (qf)   {$q_f$};  
  \path [->]  (q0)   edge [above, align=center]     node { $inf,\phi_S$\\ $x < \tau $} (qb)
		      (q0)   edge [above, align=center]  node {$inf,\phi_S$\\ $x \geq \tau$} (qf);    
\end{tikzpicture}
}
\subfigure[]{
\label{step1}
\begin{tikzpicture}[on grid, shorten <=1pt, >=stealth', auto]
  \node at (0,0)   [initial above, state] (q0)   {$q_0$};
  \node at (2.5,0)   [state] (q1)   {$q_1$};
  \node at (5,0)   [state, accepting] (qf)   {$q_f$};
  \node at (2.5,-1)   [state] (qb)   {$q_b$};
  
  \path [->] (q0)   edge [ above, align=center ]  node { $inf,\phi_S$\\ $1\leq x \leq 2$} (q1)
			      edge [ below , align=center ]  node {$inf,\phi_S$\\ $x \leq 1$} (qb)
		   (q1)   edge [ above, align=center ]  node {$inf,\phi_S$\\ $x \leq 4$} (qf);    
\end{tikzpicture}

}
\end{center}
\end{footnotesize}
\caption{The 1gDTA specifications discussed in Example \ref{ex:property} .}
\label{DTAprop}
\end{figure}
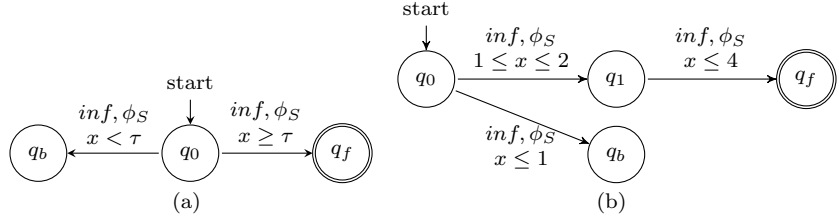

\begin{example}  \label{ex:property}
As an example, consider the agent class of the network epidemics shown in Figure \ref{SIRagent}, and the 1gDTA specification of Figure \ref{DTAprop} (a),
where the formula $\phi_S$ is true in local state $S$ and false in states $I$ and $R$. The property is satisfied  when a susceptible node is infected by an internal infection after the first $\tau$ units of time.  The sink state $q_b$ is used to discard agents being infected before $\tau$ units of time.  The use of the state formula $\phi_S$ allows us to focus only on agents that get infected, rather than also on agents that spread the contagion.
We can describe more complex properties as the 1gDTA specification of Figure \ref{DTAprop} (b). This automaton describes the fact that an agent is infected by an internal contact twice, the first infection happening between time 1 and 2, and  the second infection happening before time 4. Also here, the sink state $q_b$ is used to discard agents being infected for the first time before time 1.
\end{example}
%
%
%
%

\medskip

%

\medskip

A run $\rho$ of a 1gDTA $\dta$ is a sequence of states of $Q$, actions and times taken to change state, $q_0\xrightarrow{\alpha_0,t_0}q_1\xrightarrow{\alpha_1,t_1}\ldots q_n$, such that clock constraints are satisfied. 
A run is accepting if $q_n\in F$. 

Consider now a population model $\pop\upsize$, and focus on a single individual agent 
of class $\class$ in the population. A path $\sigma$ of length $n$ for such an agent is a sequence of the form $s_0 \xrightarrow{\alpha_0,t_0} s_1 \xrightarrow{\alpha_1,t_1} s_2 \xrightarrow{\alpha_2,t_2} \ldots s_n$, where $s_i \in \cstsp$, $t_i\in \mathbb{R}_\geq 0 $ is the time spent in the local state $s_i$, and $\alpha_i$ is the action taken at step $i$. The set of those paths will be denoted by $Path^n[\class]$ and the set of paths of finite length by $Path^*[\class]$. Given $\sigma$, we let $\tau[\sigma] = \sum_{i=0}^{|\sigma|-1} t_i$ be the total time taken to go from state $s_0$ to state $s_n$, and with $\tau_i[\sigma]$ the time taken to reach state $s_i$.
The set of paths of total duration equal to $T\in\mathbb{R}_{\geq 0}$ is denoted by $Path^T[\class]$.
Given a path $\sigma$ of length $n$, we define the run $\rho_\sigma$ of a 1gDTA $\dta$ induced by $\sigma$ to be $q_0\xrightarrow{\alpha_0,t_0}q_1\xrightarrow{\alpha_1,t_1}\ldots q_n$, where state $q_{i+1}$ is determined by the unique transition $q_i \xrightarrow{\alpha_i,\phi,c} q_{i+1}$ such that $s_i \models_{\atprop} \phi$ and  $T_{i+1}[\sigma] \models_{\calC\calC} c$. If $\rho_\sigma$ is accepting, we write $\sigma \models \dta$.

Given a 1gDTA $\dta$, we indicate it with $\dta[\phi_1,\ldots,\phi_k]$ when we want to explicitly list all the atomic propositions $\atprop$ used to build the state propositions $\calB(\atprop)$. This will be needed to define the local logic CSL-TA.

\begin{definition}[CSL-TA]
\label{def:CSLTA}
A CSL-TA formula $\Phi$ on a agent class $\class$ is defined recursively as
\[ \mathtt{true}~|~a~|~\neg\Phi~|~\Phi_1\wedge\Phi_2~|~\P^{\leq T}_{\bowtie p}\left(\dta[\Phi_1,\ldots,\Phi_k] \right),\] 
where $a$ is an atomic proposition interpreted on $\cstsp$, $T\in \mathbb{R}_{\geq 0}$, $p\in[0,1]$, $\bowtie \in \{<,\leq,\geq,>\}$, and $\dta[\Phi_1,\ldots,\Phi_k]$ is a 1gDTA with atomic formulae taken to be CLS-TA formulae $\Phi_1,\ldots,\Phi_k$.
\end{definition}
This definition is similar to \cite{cslta}, with the only difference of the use of a restricted class of DTA, and of the time bound on the probability operator $T$. The satisfaction relation is defined relatively to state $s\in\cstsp$ of an individual agent $Y(t)$ in $\pop\upsize$ of class $\class$  and an initial time $t_0$.  The only interesting case is the one involving 1gDTA specifications,  which requires a 1gDTA path property to hold with probability satisfying the bound  $\bowtie p$:
\[ s,t_0 \models \P^{\leq T}_{\bowtie p}\left(\dta[\Phi_1,\ldots,\Phi_k] \right)\ \text{iff}\ \bbP\{ \sigma\in Path^T[\class] ~|~\sigma\models \dta[\Phi_1,\ldots,\Phi_k] \} \bowtie p  \]


We turn now to introduce global properties. Here, we will consider basic properties that lift local specifications to the collective level,  looking at the fraction/ number of agents in the population that satisfy a local specification, given either as a state CSL-TA formula $\Phi$, or as a path property specified by a 1gDTA $\dta$. More specifically, we ask ourselves if the fraction/ number of agents satisfying $\dta$ (resp. $\Phi$) is included in the interval $[a,b]$, which we denote by $\dta \in [a,b]$. This is a random event, hence we need to compute its probability. In atomic global properties, we will compare this probability with a given threshold. Therefore, we have two kinds of global atomic properties:
\begin{description}
\item[Path properties:] $\gp{\gprop{\dta}{a}{b}}{}{\bowtie  p}$: the probability that the fraction/ number of agents that satisfy the local path property $\dta$ is in the interval $[a,b]$ is $\bowtie p$, for $\bowtie\in\{<,\leq,\geq,>\}$;
\item[State properties:]  $\gp{\gprop{\Phi}{a}{b}}{}{\bowtie  p}$: the probability that the fraction/ number of agents that satisfy the local state property $\Phi$ is in the interval $[a,b]$ is $\bowtie p$, for $\bowtie\in\{<,\leq,\geq,>\}$.
\end{description}
Both properties above can be checked at any starting time $t_0$.  
Atomic global properties are then combined together by boolean operators, to define more expressive queries. We therefore define a \emph{collective or global property}, as
\begin{definition}
\label{def:globalProp}
Given a population model $\pop\upsize$, a collective/ global property on $\pop\upsize$ is given by the following syntax:
\[ \Psi = \mathtt{true}~|~\gp{\gprop{\dta}{a}{b}}{\bowtie  p}~|~\gp{\gprop{\Phi}{a}{b}}{\bowtie p}~|~\neg\Psi~|~\Psi_1\wedge\Psi_2\]
\end{definition}
 
 \begin{example}
 As an example, consider again the 1gDTA property $\dta$ of Figure \ref{DTAprop}(b). Then the atomic global property $\gp{\gpropl{\dta}{\frac{N}{3}}}{}{\geq 0.8}$ specifies that, with probability at least 0.8, no more than one third of network nodes will be infected twice in the 4 time units by an internal contact, where the first infection happens in between time 1 and 2. 
\end{example}

\begin{remark} \label{rem:timeDependentDTA}
In Definition \ref{def:CSLTA}, we allow the arbitrary nesting of CSL properties within 1gDTA. By the discussion of \cite{fluidmc}, this operation requires some care when we want to apply fluid approximations to estimate probabilities. The problem is that individual agents are time-dependent non-Markovian processes (in fact, they are projections/ marginal distributions of a Markov processes, the global  model), for which the satisfaction of a CSL-TA formula involving the probability quantifier depends on the initial time at which the formula is evaluated. Hence, the satisfaction of a CSL-TA formula is a time-dependent function, while 1gDTA require time independent state formulae. This discrepancy can be reconciled by encoding this time dependency in the 1gDTA itself, using clock constraints. Hence, a state formula that is true in $s$ up to time 5 and false afterwards, will give rise to two edges in the 1gDTA, the first considering a state formula in which $s$ is true, and with an additional clock constraint $x\leq 5$, while the second corresponding to an edge with a state formula falsified by $s$, and additional clock constraint $x>5$.
%

More specifically, we consider a family $\phi_t$ of boolean state propositions $\atprop$, indexed by a time index $t\in[0,T]$, whose satisfaction value can change a finite number of times up to time $T$. This means that the interval $[0,T]$  of interest can be partitioned into a finite interval cover $[0,t_1)$, $[t_1,t_2)$,\ldots, $[t_n,T]$, such that the satisfaction of $\phi_t$ in each $[t_i,t_{i+1})$ is constant, meaning that for each state $s$ and times $t,t'\in [t_i,t_{i+1})$, then $s\models \phi_t$ if and only if $s\models \phi_{t'}$. To reduce such automata to  1gDTA, the idea is to replace the edge $q \xrightarrow{\alpha, \phi_t, c}q'$ with a collection of edges $\{q \xrightarrow{\alpha, \phi_{t_i}, c \wedge (t_i \leq x < t_i+1)}q'~|~i=0,\ldots,n\}$. This operation will be referred in the following as \emph{structural resolution of timed-varying properties}.
Once this is done, one has to check that the so-obtained DTA satisfies the determinism condition. This follows trivially if  the original DTA is explicitly deterministic.
\end{remark}

\

\bigskip

\section{Stochastic Approximations}
\label{sec:cla}

In this section, we briefly present several approaches to approximate a population model by a simpler system that allows us to keep the state space explosion under control. 
Probably the most widespread way to approximate population models is the mean field or fluid limit, which is usually accurate for large populations \cite{tutorial}. In the mesoscopic regime, i.e. for populations in the order of hundreds of individuals in which fluctuations cannot be neglected, one can rely on the linear noise approximation, treating the state space as continuous and noise as Gaussian \cite{vankampen}. As an alternative, when we are interested in moments of the population process, we can rely on a large gamma of moment closure techniques \cite{schnoerr2015}.

\paragraph{Fluid Approximation}
Given a population model $\pop\upsize = (\class, \ptrsp\upsize, \pinst\upsize)$, the Fluid Approximations provides an estimation of the stochastic dynamics of $\pop\upsize$, exact in the limit of an {\it infinite} population. In particular, we consider an infinite sequence $(\pop\upsize)_{\size \in \mathds{N}}$ of population models, all sharing the same structure, for increasing population size $\size \in \mathds{N}$ (e.g.\ the network models $(\pop\upsize_{net})_{\size \in \mathds{N}}$ with an increasing number of network nodes).
To compare the dynamics of the models in the sequence, we consider the \textit{normalised counting variables} $\hat{\X} = \frac{1}{N}\X$ (known also as \textit{population densities} or \textit{occupancy measures}, see \cite{tutorial} for further details) and we define the \textit{normalized population models} $\popn\upsize = (\class, \pntrsp\upsize,\pninst\upsize)$, obtained from $\pop\upsize$ by making the rate functions depend on the normalised variables and rescaling the initial conditions. 
For simplicity, we assume that the rate function of each transition $\tau\in\pntrsp\upsize$ satisfies the \textit{density dependent condition} $\frac{1}{\size} f_{\tau}\upsize(\pnstvec) = f_\tau(\pnstvec)$ for some Lipschitz function $f_\tau: \mathds{R}^n \longrightarrow \mathds{R}_{\geq 0}$, i.e. rates on normalised variables are independent of $N$. Also the \textit{drift} $\drift$ of $\pop\upsize$, that is the mean instantaneous change of the  normalised variables, is given by $\drift(\pnstvec) = \sum_{\tau \in \pntrsp\upsize} \textbf{v}_{\tau} f_{\tau}(\pnstvec)$ and, thus, is independent of $N$.
The unique solution\footnote{The solution exists and is unique because $F$ is Lipschitz continuous, as each $f_\tau$ is.} \[\fluid: \mathds{R}_{\geq 0} \longrightarrow \mathds{R}^n\] of the differential equation 
\begin{equation}
\label{eqn:fluid}
\frac{d\fluid(t)}{dt} = \drift(\fluid(t)),\quad  \text{given } \fluid(0) = \pninst\upsize,
\end{equation}
is the {\it Fluid Approximation} of the CTMC $\pnstvec\upsize(t)$ associated with $\popn\upsize$, and has been successfully used to describe the collective behaviour of complex systems with large populations \cite{tutorial}. The correctness of this approximation in the limit of an infinite population is guaranteed by the Kurtz Theorem \cite{kurtz,tutorial}, which states that $\sup_{t \in [0, T]}\| \pnstvec\upsize(t) - \fluid(t)\|$ converges to zero (almost surely) as $N$ goes to infinity:

\begin{theorem}
\label{th:kurtz}
Suppose $\lim_{\size \rightarrow \infty} \pninst\upsize = \textbf{x}_0$. For every \textnormal{finite} time horizon $T > 0$:
$$
\lim_{\size \rightarrow \infty} \sup_{t \in [0, T]}\left\| \pnstvec\upsize(t) - \fluid(t)\right\| = 0\qquad\text{almost surely}.
$$ 
\end{theorem}

\paragraph{Fast Simulation.}
The mean field convergence theorem can be used also to approximate the behaviour of a random individual agent in a large population model. The idea is that, in the limit of large populations, the behaviour of individual agents becomes independent, and influenced from the rest of the system only through the solution of the mean field equation. This result is known in literature as fast simulation \cite{darling,boudec,fluidmc}. 

More formally, call $Y\N(t)$ the stochastic process of a random individual agent, with state space $S$, and $\X\N(t)$ the CTMC associated with the population model. If we consider an individual agent conditional on the state $\x$ of the full population model, we can write down its infinitesimal generator matrix $Q(\x)$ as a function of $\x$. Formally this is obtained by computing the fraction of the total rate of a transition seen by a random agent  in a given state $s$, i.e. by dividing the total rate by the number of individuals in state $s$.  A more formal treatment for population models will be given in the next section. 

For the moment, observe that in a finite population, $Y\N(t)$ and $\X\N(t)$ are not independent, and to track the behaviour of an individual agent, we need to solve the full model $(Y\N(t),\X\N(t))$. However, the fast simulation theorem below proves  \cite{darling} that, in the limit of $N$ going to infinity, we can approximate the behaviour of $Y\N(t)$ by the agent $y(t)$, with state space $S$ and time-dependent infinitesimal generator matrix given by $Q( \fluid(t))$, with $ \fluid(t)$ the solution of the fluid equation presented above:
\begin{theorem} \label{th:fastsim}
For any $T<\infty$, \[\lim_{N\rightarrow \infty}P\{Y\N(t)\neq y(t),\ \text{for some } t\leq T\} = 0.\]
\end{theorem}

\paragraph{Linear Noise}
While the Fluid Approximation correctly describes the transient collective behaviour for very large populations, it is less accurate when one has to deal with a \textit{mesoscopic} system, meaning a system with a population in the order of hundreds of individuals and whose dynamics results to be intrinsically probabilistic. Indeed, the (stochastic) behaviour  becomes increasingly relevant as the size of the population decreases.  The technique of \textit{Central Limit Approximation} (CLA), also known as \textit{Linear Noise Approximation}, provides an alternative and more accurate estimation of the stochastic dynamics of mesoscopic systems. In particular, in the CLA, the probabilistic fluctuations about the average deterministic behaviour (described by the fluid limit) are approximated by a Gaussian process.

Two equivalent approaches can be followed to introduce the Central Limit Approximation: the more intuitive van Kampen's one in \cite{vankampen}, known as System Size Expansion (SSE), and a more rigorous derivation by Kurtz in \cite{kurtz}, exploiting the theory of stochastic integral equations.
In both these approaches, the idea is to focus on  the process  $\linear\upsize(t)\ :=\ \size^{\frac{1}{2}}\left(\pnstvec\upsize(t) - \fluid(t)\right)$, capturing the rescaled fluctuations of the Markov chain around the fluid limit. 
Then, by relying on convergence results for Brownian motion, one shows that $\{\linear\upsize(t)\ |\ t \in \mathds{R}\}$, for large populations $\size$, can be approximated \cite{vankampen,kurtz}  by the Gaussian process\footnote{A Gaussian process $\linear(t)$ is characterised by the fact that the joint distribution of $\linear(t_1),\ldots,\linear(t_k)$ is a multivariate normal distribution for any $t_1,\ldots,t_k$.} $ \{\linear(t) \in \mathds{R}^n\ |\ t \in \mathds{R}\}$ (\textit{independent of} $N$), whose mean $\mean [t]$ and covariance $\cov [t]$ are given by
\begin{equation}\label{mean}
\begin{cases}
\frac{\partial \mean[t]}{\partial t} = \jac_{\drift}(\fluid(t)) \mean[t]\\
\mean [0] = 0
\end{cases}
\end{equation} 
and
\begin{equation}
\begin{cases}\label{covariance}
\frac{\partial \cov[t]}{\partial t} = \jac_{\drift} (\fluid(t))\cov[t] + \cov[t]\jac^{T}_{\drift}(\fluid(t)) + \diff(\fluid(t))\\
\cov[0] = 0,\end{cases}
\end{equation} 
where $\jac_{\drift}(\fluid(t))$ denotes the Jacobian of the limit drift $\drift$ calculated along the deterministic fluid limit $\fluid: \mathds{R}_{\geq 0} \longrightarrow \mathds{R}^n$, and   $\diff(\pnstvec) = \sum_{\tau \in \pntrsp\upsize} \textbf{v}_{\tau}\textbf{v}_{\tau}^T f_{\tau}(\pnstvec)$ is called the \emph{diffusion} term. 


The nature of the approximation of $\linear\upsize(t)$ by $\linear(t)$ is captured in the following theorem~\cite{kurtz}.

\begin{theorem}
\label{th:central}
Let $\linear(t)$ be the Gaussian process with mean (\ref{mean}) and covariance (\ref{covariance}) and $\linear\upsize(t)$ be the random variable given by $\linear\upsize(t)\ :=\ \size^{\frac{1}{2}}\left(\pnstvec\upsize(t) - \fluid(t)\right)$. Assume that $\lim_{\size \rightarrow \infty} \linear\upsize(0) = \linear(0)$. Then, $\linear\upsize(t)$ converges in distribution to $\linear(t)$.\footnote{Formally,  $\{\linear\upsize(t)\}_{t\in \mathds{R}_{\geq 0}} \Rightarrow \{\linear(t)\}_{t\in \mathds{R}_{\geq 0}}$, i.e. the convergence in distribution is of $\linear\upsize$ to $\linear$, seen as random variables taking values in the space of cadlag functions from $\mathds{R}^n$ to $\mathds{R}$. }
\end{theorem}
The  \textit{Central Limit Approximation} then approximates the normalized CTMC $\pnstvec\upsize(t) = \fluid(t) + \size^{-\frac{1}{2}} \linear\upsize(t)$ associated with $\popn\upsize$ by the stochastic process
\begin{equation}\label{vanassumption}
\fluid(t) + \size^{-\frac{1}{2}} \linear(t).
\end{equation}
Theorem \ref{th:central} guarantees its asymptotic correctness in the limit of an infinite population.

In the derivation of Van Kampen \cite{vankampen}, the idea is to start from the master equation and introduce an expansion around a small parameter given by the inverse of the system size. In this way, truncating at the lowest order, one obtains that fluctuations obey a linear Fokker-Plank equation, whose solution is the Gaussian Process described by the mean and covariance function introduced above. The advantage of this derivation is that one can keep higher orders in the expansion \cite{grima2010}, obtaining a non-linear Fokker-Plank, which cannot be solved analytically, but which can be either integrated numerically, or from which equations for moments can be extracted. These equations have higher-order correction terms depending on system-size, which vanish in the thermodynamic limit, collapsing to mean field. We do not present the derivation here in detail, but refer the interested reader to the appropriate papers. It is worth mentioning that this approach has been implemented in the tool iNA \cite{ina} and in the matlab toolbox CERENA \cite{CERENA}. 

\paragraph{Moment Closure}
The class of approximation techniques known as moment closure \cite{schnoerr2016} is a viable  alternative to mean field, linear noise, or SSE, when the interest is on the moments of the population process for a finite population rather than on the limit behaviour for large population sizes. These methods start from a general ODE for the moments of a stochastic process, known as Dynkin formula \cite{kallenberg}. If $h(\x)$ is any sufficiently smooth function with domain $\bbR^n$, then 
\[ \frac{d}{dt}\bbE_t[h(\X\upsize(t))] =   \sum_{\tau \in \pntrsp\upsize} \bbE_{t}[ (h(\X\upsize(t) + \textbf{v}_{\tau}) - h(\X\upsize(t)) )f\upsize_{\tau}(\X\upsize(t)) ]. \]
In the formula above, $\tau \in \pntrsp\upsize$ are the transitions of a population model, each with rate $f\upsize_{\tau}(\x)$ and with update vector $\textbf{v}_{\tau}$, see Section \ref{sec:popmod}. 
Starting from this formula, one can easily obtain the exact ODEs for (non-centered) moments, by using a suitable polynomial expression for $h$. For instance, for $h(\x) = x_i$ one can obtain the mean of population  $i$, with $h(\x) = x_i^2$ one obtains the second moment of population $i$, from which variance can be computed as $\bbE[X_i^2] - \bbE[X_i]^2$, and so on.
It is useful to see what happens when $h$ is the vector valued identity, $h(\x) = \x$, giving the mean for all variables. In this case, the ODE above becomes 
\[ \frac{d}{dt}\bbE_t[\X\upsize(t)] =   \sum_{\tau \in \pntrsp\upsize} \textbf{v}_{\tau} \bbE_{t}[ f\upsize_{\tau}(\X\upsize(t)) ], \]
hence the right hand side depends on the expectation of the rate functions with respect to the state of the Markov process $\X\upsize(t)$ at time $t$.  If all rate functions are linear, then the previous equation is closed, i.e. depends only on the mean $\bbE[\X\upsize(t)]$. However, when  the rate functions are non-linear polynomials, like in the epidemic example, or more complex non-linear functions, then this is no more the case. For example, in the epidemic model of Section , we can observe how the infection rate is $k_iX_S X_I$, hence giving rise to a term in the ODE for the mean equal to $k_i\bbE[X_S X_I]$, i.e. the covariance between $X_S$ and $X_I$. This is the \emph{leit motiv} of non-linearity: differential equation for a moment of order $m$ will depend on moments of higher order, thus giving rise to an infinite dimensional ODE system. 
The whole business of moment closure is to find clever ways to truncate these infinite dimensional ODEs to obtain a finite dimensional set of equations up to order $m$ to be solved by numerical integration.
In literature, there are different techniques to close the moment equations. Typically they impose some condition that is satisfied by a family of probability distributions, from moments of order $m$ on (e.g. normal, log-normal, low dispersion), but they may also try to match the derivatives to those of the true equations, or use other ideas, see \cite{schnoerr2014,schnoerr2016,CERENA} for a presentation of different moment closure strategies.
Typically, the higher the order $m$ of truncation, the more accurate will be the estimate of lower order moments, like the mean. This is not always true, as the accuracy of moment closures depends on the system under consideration in complex ways, see \cite{schnoerr2014} for a discussion in this sense. 
In the paper, we make use of the low dispersion moment closure, which can be obtained by setting the centred moments to zero from order $m$ on, see  \cite{schnoerr2016,CERENA} .\\

%

\section{Model Checking CSL-TA Properties on Individual Agents}
\label{sec:checkCSLTA}

In this section, we show how to check CSL-TA formulae on individual agents. The challenging task is that of computing path properties for 1gDTAs, restricting to constant atomic propositions (in the light of Remark \ref{rem:timeDependentDTA}). In this paper, we will consider efficient approximations of the path probability based on fluid approximations (Section \ref{sec:singlePath}). 
The computation of path probabilities starts from the construction of an enriched agent, obtained by a product construction between the agent model with the automata of the properties (Section \ref{sec:singleSynch}). Then, we show how to compute path probabilities from a fixed initial time (Section \ref{sec:singlePathFixed}), and then as a function of the initial time (Section \ref{sec:singlePathVariable}). 
The next step is to provide an approximate model checking algorithm for CSL-TA (Section \ref{sec:singleMC}), discussing its decidability and the convergence to the exact value for large populations (Section \ref{sec:singleConvergence}). 
Finally, we show how to improve the accuracy of the approximation for finite populations, exploiting higher order moment closure techniques (Section \ref{sec:singleHO}).

\subsection{Computing path probabilities} \label{sec:singlePath}

\subsubsection{Synchronisation of agents and 1gDTAs.} \label{sec:singleSynch}

The first step in the computation of path probabilities  is to synchronize the agent and the property, constructing an extended Markov population model in which the state space of each agent is combined with the specific path property we are observing.
The main  difficulty in this procedure is the presence of time constraints in the path property specification. However, thanks to the restriction to a single global clock, we can partition the time interval of interest into a finite set of subintervals or regions, within which no clock constraint changes status. Thus, in each region, we can remove the clock constraints, deleting all the edges that cannot fire because their clock constraints are false. In this way, we generate a sequence of Deterministic Finite Automata (DFA), that are then combined with the local model $\class$ by a standard product of automata. 

Let $\class = (\cstsp, \ctrsp)$ be an agent class,  $\dta = (\lbset, \atprop, \gstsp, \ginst, \gfinst, \gtr)$ be a local path property, and $T> 0$ be the time horizon.

\vspace{0.2cm}

\noindent \textbf{First step: enforcing uniqueness of transition labels}. We define a new agent class $\bar{\class} = (\cstsp, \bar{\ctrsp})$ by renaming the local transitions in $\ctrsp$ to make their label unique.  This allows us to remove edge formulae in $\dta$, simplifying the product construction. In particular, if there exist $s_1 \xrightarrow{\loclab{}} s'_1, \ldots, s_m \xrightarrow{\loclab{}} s'_m \in \ctrsp$ having the same label $\loclab{}$, we rename them by $\loclab{s_1}, \ldots, \loclab{s_m}$, obtaining $s_1 \xrightarrow{\loclab{s_1}} s'_1, \ldots, s_m \xrightarrow{\loclab{s_m}} s'_m \in \bar{\ctrsp}$. The 1gDTA $\gdta$ is updated accordingly, by substituting each edge $q\xrightarrow{\alpha,\phi,c}q'$ with the set of edges $q\xrightarrow{\alpha_{s_i},\phi,c}q'$, for $i=1,\ldots,m$. We call $\bar{\lbset}$ the label set of $\bar{\class}$.

\vspace{0.2cm}

\noindent  \textbf{Second step: removal of state conditions}. 
We remove from the edge relation of $\gdta$ all the edges $q\xrightarrow{\alpha_{s_i},\phi,c}q'$ such that $s_i\not\models_{\atprop} \phi$, where $s_i$ is the source state  of the (now unique) transition of $\bar{\class}$ labeled by $\alpha_{s_i}$. At this point, the information carried by state propositions becomes redundant, thus we drop them, writing 
$q\xrightarrow{\alpha_{s_i},c}q'$ in place of $q\xrightarrow{\alpha_{s_i},\phi,c}q'$.

\vspace{0.2cm}

\noindent \textbf{Third step: removal of clock constraints}. Let $t_1, \ldots, t_{k}$ be the ordered sequence of constants (smaller than $T$) appearing in the clock constraints of the edges of $\gdta$. We extend this sequence by letting  $t_0 = 0$ and $t_{k+1} = T$. Let $I_j = [t_{j-1}, t_{j}]$, $j = 1, \ldots, {k+1}$, be the $j$-th sub-interval of $[0,T]$ identified by such a sequence. For each $I_j$,  we define a Deterministic Finite Automaton (DFA), $\dta{j} = (\lbset, \gstsp, \ginst, \gfinst, \tr{j}{})$, whose edge relation $\tr{j}{}$ is obtained from that of $\gdta$ by selecting only the edges for which the clock constraints are satisfied in $I_j$, and dropping the clock constraint. Hence, from $q\xrightarrow{\alpha_{s_i},c}q'$  such that $\eta(x)\models_{\calC\calC} c$ whenever $\eta(x)\in (t_{j-1},t_{j})$, we obtain the DFA edge $(q,\alpha_{s_i},q')\in\tr{j}{}$, denoted also by $q\tr{j}{\alpha_{s_i}} q'$. 
%

\vspace{0.2cm}

\noindent \textbf{Fourth step: synchronization}. To keep track of the behaviour of the agents with respect to the property specified by $\gdta$, we synchronize the agent class $\bar{\class} = (S, \bar{E})$ with each DFA $\dta{j}$ 
through the standard product of automata. The sequence of deterministic automata obtained in this procedure is called the \textit{agent class associated with the local property} $\gdta$.

\begin{definition}[Agent class associated with the local property $\gdta$]
\label{def:synchAgentClass}
The agent class $\apro$ associated with the local property $\gdta$ is the sequence $\apro = (\aca{1}, \ldots,\\ \aca{k+1})$ of deterministic automata $\aca{j} = (\acstsp, \actrsp_{j})$, $j = 1, \ldots, k+1$, where $\acstsp = \cstsp \times \gstsp$ is the state space and $\actrsp_{j}$ is the set of local transitions $\loctr{i}^{j} = (s, q) \xrightarrow{\loclab{s}} (s', q')$, such that $s\xrightarrow{\loclab{s}} s'$ is a local transition in $\bar{\class}$ and $q\xrightarrow{\loclab{s}} q'$ is an edge in $\dta{j}$.
%
\end{definition}
%
%
%
%

\begin{example} \label{ex:syncA_DTA}
As an example, in Figure \ref{fig:sync}, we show the synchronisation steps of the SIR automata described in Figure \ref{SIRagent}, and the 1gDTA specification of Figure \ref{DTAprop} (a). After the first step, Figure \ref{fig:sync} (a), the new agent class $\bar{\class}_{node} = (\cstsp_{node}, \bar{\ctrsp}_{node})$ of the network node has local transitions $\bar{\ctrsp}_{node} =\{S\xrightarrow{inf_S} I, S\xrightarrow{ext} I, I\xrightarrow{inf_I} I, I\xrightarrow{patch_1} R, S\xrightarrow{patch_0} R, R\xrightarrow{loss} S \}$. The 1gDTA $\gdta$ is updated accordingly, by substituting the edge $q_0\xrightarrow{inf,\phi_S, c}q_i$ with the set of edges $q_0 \xrightarrow{inf_S,\phi_S,c}q'$, for $i= b,f$.   We remove then, Figure \ref{fig:sync} (b), the redundant state conditions of $\gdta$.
In the third step, Figure \ref{fig:sync} (c), we remove the clock constraints. For the considered 1gDTA we have two time intervals $I_1 = [0, \tau]$ and $I_2 = [\tau ,T]$. We define then two DFAs: $\mathscr{D}_{[0, \tau ]} $ and $\mathscr{D}_{[\tau, T ]} $. Finally, Figure \ref{fig:sync} (d), we synchronise the agent class $\bar{\class}_{node}$ with each DFA.

\begin{figure}[H]
\begin{center}
%
%
%
\subfigure[Enforcing uniqueness of transition labels]{
\label{step1}
\begin{tikzpicture}[on grid, shorten <=1pt, >=stealth', auto]
  \node at (0,0)   [state] (I)   {{\bf I}};
  \node at (3,0)   [state] (R)   {{\bf R}};
  \node at (1.5,2) [state] (S)   {{\bf S}};
  
  \path [->] (I)  edge [loop left, left]     node {\textcolor{red}{$inf_I$}}    ()
                  edge [bend right, below]      node {$patch_1$}    (R)
             (R)  edge [bend right, right]       node {$loss$}   (S)
             (S)  edge [bend right, left]      node {\textcolor{red}{$inf_S$}}    (I)
                  edge [right]                  node {$ext$} (I)
                  edge [left, very near end]       node {$patch_0$} (R);
\end{tikzpicture}
\hspace{5mm}
\begin{tikzpicture}[on grid, shorten <=1pt, >=stealth, auto]
  \node at (2,0)   [initial above,state] (q0)   {$q_0$};
  \node at (0,0)   [state] (qb)   {$q_b$};
  \node at (4,0)   [state, accepting] (qf)   {$q_f$};  
  \path [->]  (q0)   edge [above, align=center]     node { \textcolor{red}{$inf_S$}, $\phi_S$\\ $x < \tau $} (qb)
		      (q0)   edge [above, align=center]  node {\textcolor{red}{$inf_S$}, $\phi_S$\\ $x \geq \tau$} (qf);    
\end{tikzpicture} 
}

\vspace{3mm}
\subfigure[Removal of state conditions]{
\label{step2}
%
\begin{tikzpicture}[on grid, shorten <=1pt, >=stealth, auto]
  \node at (2,0)   [initial above,state] (q0)   {$q_0$};
  \node at (0,0)   [state] (q1)   {$q_b$};
  \node at (4,0)   [state, accepting] (qf)   {$q_f$};  
  \path [->]  (q0)   edge [above, align=center]     node { $inf_S$, \textcolor{red}{$\cancel{ \phi_S}$}\\ $x < \tau $} (qb)
		      (q0)   edge [above, align=center]  node {$inf_S$, \textcolor{red}{$\cancel{\phi_S}$}\\ $x \geq \tau$} (qf);    
\end{tikzpicture}
}

\vspace{3mm}
\subfigure[Removal of clock constraints]{
\label{step3}
%
$\mathscr{D}_{\textcolor{red}{[0, \tau ]}} =$ 
\begin{tikzpicture}[on grid, shorten <=1pt, >=stealth, auto]
  \node at (2,0)   [initial above,state] (q0)   {$q_0$};
  \node at (0,0)   [state] (qb)   {$q_b$};
  \path [->]  (q0)   edge [above, align=center]     node { $inf_S$} (qb);
\end{tikzpicture}
\hspace{5mm}
$\mathscr{D}_{\textcolor{red}{[\tau, T]}} =$ \begin{tikzpicture}[on grid, shorten <=1pt, >=stealth, auto]
  \node at (2,0)   [initial above,state] (q0)   {$q_0$};
  \node at (4,0)   [state, accepting] (qf)   {$q_f$};  
 \path [->] 
		      (q0)   edge [above, align=center]  node {$inf_S$} (qf);    
\end{tikzpicture}
}

\vspace{3mm}
\subfigure{
\label{step4}
$\mathscr{P}_{\textcolor{red}{[0, \tau]}} =$
\begin{footnotesize} 
\begin{tikzpicture}[on grid, shorten <=1pt, >=stealth, auto]
  \node at (0,0)       [state] (s00)  {${\bm I_{0}}$};
  \node at (2,0)       [state] (s20)  {${\bm R_{0}}$};
  \node at (1,2)     [state] (s10)  {${\bm S_{0}}$};
  \node at (4,0)       [state] (s01)  {${\bm I_{b}}$};
  \node at (6,0)       [state] (s21)  {${\bm R_{b}}$};
  \node at (5,2)     [state] (s11)  {${\bm S_{b}}$};

  \path [->] (s00)  edge [loop left, below]       node {$inf_{I,0}$}      ()
                    edge [bend right, below]      node {$patch_{1,0}$}    (s20)
             (s20)  edge [bend right, right, very near start]      node {$loss_{0}$}       (s10)
             (s10)  edge [bend left, below]      node {\textcolor{red}{$inf_{S,0}\phantom{.....}$}}      (s01)
                    edge [right]                  node {$ext_{0}$}        (s00)
                    edge [left, very near end]    node {$patch_{0,0}$}    (s20)
             (s01)  edge [loop left, below]       node {$inf_{I,b}$}      ()
                    edge [bend right, below]      node {$patch_{1,b}$}    (s21)
             (s21)  edge [bend right, right, very near start]      node {$loss_b$}       (s11)
             (s11)  edge [bend right, left, very near start]       node {$inf_{S,b}$}      (s01)
                    edge [right]                  node {$ext_b$}        (s01)
                    edge [left, very near end]    node {$patch_{0,b}$}    (s21);
\end{tikzpicture}
\end{footnotesize} 
}
\vspace{3mm}
\subfigure[Synchronisation]{
\label{step4}
$\mathscr{P}_{\textcolor{red}{[\tau, T]}} =$
\begin{footnotesize} 
\begin{tikzpicture}[on grid, shorten <=1pt, >=stealth, auto]
  \node at (0,0)       [state] (s00)  {${\bm I_{0}}$};
  \node at (2,0)       [state] (s20)  {${\bm R_{0}}$};
  \node at (1,2)     [state] (s10)  {${\bm S_{0}}$};
  \node at (4,0)       [state, accepting] (s01)  {${\bm I_{f}}$};
  \node at (6,0)       [state, accepting] (s21)  {${\bm R_{f}}$};
  \node at (5,2)     [state, accepting] (s11)  {${\bm S_{f}}$};

  \path [->] (s00)  edge [loop left, below]       node {$inf_{I,0}$}      ()
                    edge [bend right, below]      node {$patch_{1,0}$}    (s20)
             (s20)  edge [bend right, right, very near start]      node {$loss_{0}$}       (s10)
             (s10)  edge [bend left, below]      node {\textcolor{red}{$inf_{S,0}\phantom{.....}$}}      (s01)
                    edge [right]                  node {$ext_{0}$}        (s00)
                    edge [left, very near end]    node {$patch_{0,0}$}    (s20)
             (s01)  edge [loop left, below]       node {$inf_{I,f}$}      ()
                    edge [bend right, below]      node {$patch_{1,f}$}    (s21)
             (s21)  edge [bend right, right, very near start]      node {$loss_f$}       (s11)
             (s11)  edge [bend right, left, very near start]       node {$inf_{S,f}$}      (s01)
                    edge [right]                  node {$ext_f$}        (s01)
                    edge [left, very near end]    node {$patch_{0,f}$}    (s21);
\end{tikzpicture}
\end{footnotesize}
}
\end{center}
\caption{ Synchronisation steps of the SIR automata described in Figure \ref{SIRagent}, and the 1gDTA specification of Figure \ref{DTAprop} (a).}
\label{fig:sync}
\end{figure}
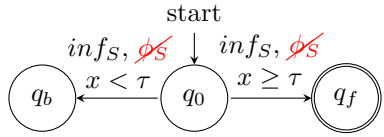
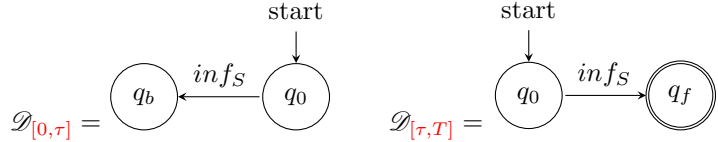
 \end{example}

\subsubsection{The stochastic model of the single agent.} \label{sec:agentModel}
The synchronisation of an agent with a property enables us to monitor if and when the agent satisfies it. To progress further into verification, we need to tweak the system model so that one agent is tagged and monitored during model execution. The idea to do this is simple: we couple the agent with the population model, and each time a global transition fires which can change the current state of the agent,  we choose whether to update the tagged agent or another untagged one. The best way to formalize this is to define the infinitesimal generator of an individual agent, conditional on the state of the population model. In turn, to specify this we just need to specify for each local transition of the agent class of Definition \ref{def:synchAgentClass} the rate at which an individual agent will see this transition happen, given the current state $\X\upsize(t)$ of the global model. To fix the notation, let us denote by $\tY\upsize(t)$ the state of the tagged agent at time t. 

Consider now a transition $\loctr{i}^{j} = (s, q) \xrightarrow{\loclab{s}} (s', q')$ of the agent of class $\apro_j$,\footnote{We consider the behaviour in a single interval $I_j$ of $\apro$.} and let $\tau$ be a global transition  of the population model such that $s \xrightarrow{\loclab{}} s'$ belongs to its synchronisation set $\syncset{\tau}$. Let $f\upsize_{\tau}: \mathds{R}^n \longrightarrow \mathds{R}_{\geq 0}$ the rate function of the transition. Furthermore, let $m_{\tau}$ be the multiplicity of  $s \xrightarrow{\loclab{}} s'$ in $\syncset{\tau}$. Then, the fraction of rate of $\tau$ seen by the individual agent can be computed by dividing the global rate by the number of agents in state $s$, and correcting for the multiplicity  $m_{\tau}$, as shown in the following

\begin{proposition} \label{prop:individualRates}
The rate of transition $(s, q) \xrightarrow{\loclab{s}} (s', q')$  of an individual agent due to global transition $\tau$, given that the population model is in state $\X\upsize(t) = \x$, is
\[ g\upsize_{\tau}((s,q),(s',q')) = \frac{m_\tau}{x_s}  f\upsize_{\tau}(\x).\]
\end{proposition}


We can now easily define the infinitesimal generator of an individual agent  of class $\apro_j$, conditional on the  population model being in state $\X\upsize(t) = \x$ as the matrix $Q_j\upsize(\x)$ such that
\[Q\upsize_{j,(s,q),(s',q')}(\x)  =   \sum_{(s, q) \xrightarrow{\loclab{s}} (s', q')\in  \actrsp_j } \sum_{\tau| s \xrightarrow{\loclab{}} s' \in\syncset{\tau}}    g\upsize_{\tau}((s,q),(s',q')).  \]
Furthermore, as customary, let $Q\upsize_{j,(s,q),(s,q)}(\x)= - \sum_{(s,q) \neq (s',q')} Q\upsize_{j,(s,q),(s,q)}(\x)$.

\begin{example} \label{ex:Q1agent}
Let us consider the synchronisation of a SIR agent  and the 1gDTA specification described in the previous subsection (Example \ref{ex:syncA_DTA}). We couple this agent with the population model $\pop\upsize_{net} = (\class_{node},  \ptrsp\upsize, \pinst\upsize)$, that has population variables $\X = (X_S, X_I, X_R)$. We can define the infinitesimal generator of the individual agent of class $\mathscr{P}_{[0, \tau]}$ and $\mathscr{P}_{[\tau, T]} $ conditional on the  population model being in state $\X\upsize(t) = (x_S, x_I, x_R)$.
For example, to change its state from $S_0$ to $I_b$ the automata can execute the local transition $S_0 \xrightarrow{inf_{S,0}} I_b$ that belongs to the synchronisation set of the global transition $\tau_{inf}$, with multiplicity $m_{\tau_{inf}}=1$. The global transition has rate function $f\upsize_{inf}(\x) =  \frac{1}{N} \kappa_{inf} x_S x_I$. The rate of the individual agent is then equal to $g\upsize_{\tau_{inf}}(S_0,I_b) =  \frac{1}{x_S}\frac{\kappa_{inf}}{N} x_S x_I = \frac{\kappa_{inf}}{N} x_I$, where $\x = (x_S, x_I, x_R)$ is the density of the population model, which can be computed at a given time by the fluid approximation. This is the only transition that allows to change state from $S_0$ to $I_b$ then  $ Q\upsize_{[0,\tau],(S_0,I_b)}(\x) =\frac{\kappa_{inf}}{N} x_I$. In a similar way we can compute the other values of the generator matrix: $Q\upsize_{[0,\tau],(S_0,I_0)}(\x)=\kappa_{ext}$, $Q\upsize_{[0,\tau],(S_0,R_0)}(\x)=\kappa_{patch_0}$. 
$ Q\upsize_{[0,\tau],(S_0,S_0)}(\x) =   - \kappa_{ext} - \frac{\kappa_{inf}}{N} x_I-\kappa_{patch_0}$. $Q_{[\tau,T]}(\x)$ is equal to $Q_{[0, \tau]}(\x)$ except for $Q_{[T,\tau], (S_0,I_b)}(\x)=0$ and $Q_{[T,\tau], (S_0,I_f)}(\x)= \frac{\kappa_{inf}}{N} x_I$. 
\end{example}

\subsubsection{Computing path probabilities for a fixed initial time} \label{sec:singlePathFixed}
Verifying the property $\dta$ on an individual agent requires to compute the path probability of the set of paths that satisfy it. This can be done by synchronising the agent with the property, as in Definition \ref{def:synchAgentClass}, and then computing the probability at the time horizon $T$ of being in an accepting state. This is sufficient because of the absorption property of final states in a 1gDTA (condition 2 of Definition \ref{def:1gDTA}), which guarantees that whenever an agent enters a final state of the 1gDTA, it will never leave it, i.e. that the second component of a state $(s,q_f)$,  $q_f\in \gfinst$, will never change.
Let the individual agent $Y\N(t)$ be in state $s_0$ at the initial time $t_0$. Then the synchronised agent will start from state $(s_0,q_0)$, and 
\begin{equation} 
\label{eq:prob}
P(s_0,t_0 \models \dta) = P( Y\N,s_0,t_0 \models \dta ) = \sum_{(s,q)|q\in F} P(\tY\N(t_0+T) = (s,q)). 
 \end{equation}

The problem with the formula above is that to compute the probabilities of $\tY\N$ one needs to solve the joint process $(\tY\N(t),\X\N(t))$, as the rates of  $\tY\N$ depend on the state of the global model.  To speed up this computation, the idea is to plug in an approximation. The simplest choice, which is typically working well for moderate to large population sizes, is to rely on the fast simulation, approximating  $\tY\N$  by $\hY(t)$, the individual agent model with time-dependent rates, plugging in $Q_j$ the solution $\x(t)$ of the mean field equation for the global model: $Q_j = Q_j(\x(t))$. This is the idea pursued in \cite{fluidmc}, which gives a speedup of many orders of magnitude. 

In this section, we proceed along this direction, but follow a different derivation which makes easier to correct the model for finite size effects, using higher order moment closure techniques.  Consider the joint distribution $P(\tY\N(t),\X\N(t))$ and write it as $P(\tY\N(t)|\X\N(t))P(\X\N(t))$. We now plug in the crucial approximation, which is a consequence of the fast simulation theorem: we assume $\tY\N(t)$ and $\X\N(t))$ to be independent. This guarantees that $\tY\N(t)$ is a Markov process, so that we can derive the forward Kolmogorov equations for the marginal process $\tY\N(t)$, as 
\begin{equation} \label{eq:Keq} 
 \frac{d}{dt} P_j(t|t_0) =  P_j(t|t_0) \bbE_{\X\N(t)}[Q_j(\X\N(t))] 
 \end{equation} 
i.e. by marginalising over the global population model. Here $P_j(t|t_0) $ is a matrix of transition probabilities: $P_j(t|t_0) [(s,q),(s',q')]$ is the probability of being in state $(s',q')$ at time $t$, starting from state $(s,q)$ at time $t_0$.

Now, the fast simulation regime introduces the further approximation
\[ \bbE_{\X\N(t)}[Q_j(\X\N(t))]  \approx Q_j(\bbE[\X\N(t)]) \approx Q_j(\x(t))\]
where we made a first order approximation of $\bbE_{\X\N(t)}[Q_j(\X\N(t))]$  by Taylor expanding it around the mean $\bbE[\X\N(t)]$, and then approximated this mean at first order with the solution of the mean field equation: $\bbE[\X\N(t)]\approx\x(t)$ \cite{vankampen,gardiner,bortolussi2008}.

Notice that, even at first order, we obtain a time-inhomogeneous model for the individual agents, with rates modulated by the average behaviour of the full process. Obviously, there is no need to stop at first order, and we can introduce higher order approximations of the average $\bbE_{\X\N(t)}[Q_j(\X\N(t))] $, by relying on moment closure approximation. We will investigate this direction in Section \ref{sec:singleHO}.

To compute the probability of the property $\dta$, we need now to take into account the structure of clock constraints. The idea is that we can apply the approximation discussed above to each synchronised model $\apro_j$, and then combining the so obtained probabilities by multiplying the probability transition matrix. More specifically, call $P_j(t_i|t_{k})$ the probability transition matrix of $\apro_j$, computed by solving the approximate Kolmogorov equations \eqref{eq:Keq}. 
Let $t_1,\ldots,t_N$ be the clock constraints. Then we define 
\begin{equation} \label{eqn:fullProbIndividual}
 P(t_N|t_0) = P_1(t_1|t_0)P_2(t_2|t_1)\cdots P_N(t_N|t_{N-1})  
 \end{equation}
and then let $P(\tY\N(t_0+T) = (s,q)) = P(t_N|t_0)[(s_0,q_0),(s,q)]$  in equation \eqref{eq:prob}. The satisfaction probability $P(s_0,t_0 \models \dta)$ can now be calculated according to equation  \eqref{eq:prob}.
%

\begin{example} \label{ex:st1agentfixTime}
In the previous subsection (Example \ref{ex:Q1agent}), we couple the automata  $\mathscr{P}_{[0, \tau]}$ and $\mathscr{P}_{[\tau, T]} $ with the population model $\pop\upsize_{net} = (\class_{node},  \ptrsp\upsize, \pinst\upsize)$ and we compute their infinitesimal generator matrix conditional on the  population model being in state $\X\upsize(t) = (x_S, x_I, x_R)$. Let us denoted by $Y\upsize(t)$ the state of the tagged agent at time t and suppose it is in state $S$ at time $t_0
=0$, The synchronised agent will start from state $S_0$. We want to compute $P( Y\N,S,t_0 \models \dta ) =  P(\tY\N(t_0+T) = S_f) + P(\tY\N(t_0+T) = I_f) +P(\tY\N(t_0+T) = R_f).$ To do that we approximate $\tY\N$ by $\hY(t)$ computing the infinitesimal generator matrix $Q_{[0,\tau]}$ and $Q_{[\tau,T]}$ and we integrate the approximate Kolmogorov equations: 
\[ \frac{d}{dt} P_{[0,\tau]}(t|0) = P_{[0,\tau]}(t|0)  Q_{[0,\tau]}(\x(t)) \qquad\qquad  \frac{d}{dt} P_{[\tau, T]}(t|\tau) = P_{[\tau,T]}(t|\tau)  Q_{[\tau,T]}(\x(t)) .\] 
Note that we have only one clock constraint $\tau$.
We can compute then $ P(T|0)= P_{[0,\tau]}(\tau|0) P_{[\tau, T]}(T|\tau)  $. The total satisfaction probability is equal to $P( Y\N,S_0,t_0 \models \dta ) = P(T|0)[S_0,S_f ]+ P(T|0)[S_0,I_f]+ P(T|0)[S_0,R_f]$.
In Figure \ref{1agfixtimefluid}, we report a comparison of the fast simulation (FS) described above with the statistical estimation (using the Gillespie algorithm, SSA with 10000 runs) of the path probabilities, as function of the time horizon $T$, computed for different values of population size $N$ (20,50,100).
In Table \ref{table:error1agfixtime}, we report the average computational cost of SSA, the computational cost FS and the relative SpeedUp (FScost/SSA cost), the  mean  and  maximum absolute and relative errors of the fast simulation in $[0,T]$. We also report the error at the final time of the simulation, when the probability has stabilised to its limit value. It can be seen that both the average and the maximum errors decrease with $N$, as expected, and are already quite small for $N=50$ (for the first property, the maximum difference in the path probability for all runs is of the order of 0.06, while the average error is 0.003). For $N=100$, the FS is practically indistinguishable from the (estimated) true probability. Moreover, the solution of the ODE system is computationally independent of $N$, and also much faster, as can be seen from the computation costs, than the simulation based method.


\begin{figure}[h]
\begin{center}
\includegraphics[width=.90\textwidth]{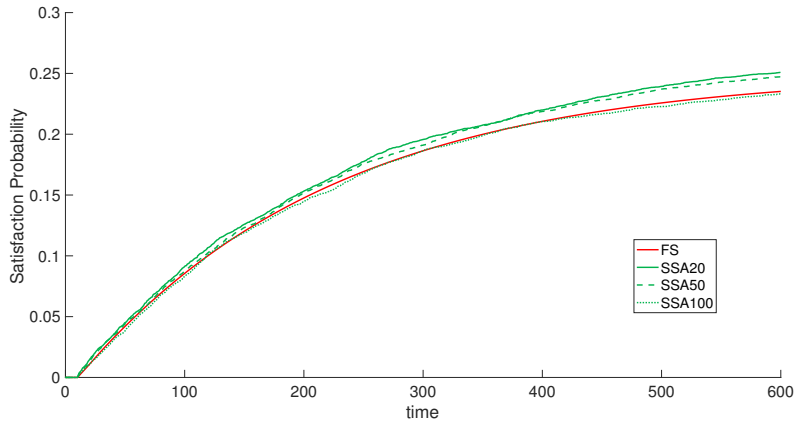}
\end{center}
\caption{Comparison of the fast simulation  (FS) and a statistical estimate (using the Gillespie algorithm, SSA with 10000 runs) of the path probabilities of the 1gDTA property of Figure \ref{DTAprop} (a) as function of the time horizon $T$, computed on the network epidemic model for different values of the population size $N$ (20,50,100). Parameters of the model are $\kappa_{inf}=1$, $\kappa_{patch_1}=0.08$, $\kappa_{loss}=0.01$, $\kappa_{ext}=0.1$, $\kappa_{patch_0}=0.001$.}
\label{1agfixtimefluid}
\end{figure}

\begin{table}[!t]
\begin{center}
\begin{small}
\begin{tabular}{|c|c|c|c|c|c|c|c|c|c|}
\hline
$N$   & SSAcost & FScost & Speedup & max(err) & $\bbE$[err] & err($T$)  & $\bbE$[Relerr] & Relerr($T$)  \\ 
\hline
    20    & 68.870 & 0.2040& 337.598  & 0.0159 &    0.0086  &  0.0158&     0.0579 &  0.0631\\
    \hline
    50    &   77.591 & 0.2040 & 380.350   & 0.0121 &    0.0062 &    0.0120 &   0.0406 &    0.0487   \\
    \hline
 100    & 97.490  & 0.2040 & 477.892  & 0.0045   &  0.0017  &  0.0020 &   0.0166  &  0.0084 \\
 \hline
  200  & 103.598 &0.2040 & 507.833 &    0.0044   & 0.0017 &    0.0018&     0.0147 &    0.0077\\
  \hline
  500   & 119.3612 &0.2040 & 585.104   &0.0041 &    0.0015 &    0.0008  &    0.0214 &    0.0033  \\
\hline
\end{tabular}
\end{small}
\end{center}

\caption{Computational cost of the statistical estimation (SSA) for 10000 runs  and of the Fast Simulation (FScost), the relative SpeedUp (FScost/SSAcost) and the errors obtained by the Fast Simulation: maximum and average absolute error  (max([er]), $\bbE$[er]) and relative error (Relerr($T$), $\bbE$[Relerr]) with respect to time, and error at the final time horizon $T$ (err($T$). Data is shown as a function of the network size $N$. }
\label{table:error1agfixtime}
\end{table}
\end{example}

\subsubsection{Computing path probabilities as a function of the initial time} \label{sec:singlePathVariable}
In the previous section we showed how to compute the satisfaction probability of a path formula for a fixed initial time, in the approximate single agent model. As the rates of the individual will depend on the global system through the expected values of some functions of the global variables, the individual agent is a time-dependent CTMC, hence the same property evaluated at different initial times can in principle have different probability values. In order to properly deal with nesting in the  model checking algorithm for the CSL-TA logic, following the approach of \cite{fluidmc}, we need to compute the path probability as a function of the initial time. 

We can apply a similar approach as in \cite{fluidmc}, which we quickly recall here. 
Consider the probability transition matrix $P(t_0+T|t_0)$. Fixing the time horizon $T$, we need to compute it as a function of $t_0$. To achieve that, we can combine the forward and backward Kolmogorov equations:
\[ \frac{\partial}{\partial t}  P(t|s) = P(t|s)Q(t)\qquad\qquad  \frac{\partial}{\partial s}  P(t|s) = -Q(s)P(t|s) .\]  obtaining the following ODE for $P(t_0+T|t_0)$:
 \begin{equation} 
 \label{eq:forwbackKE}
   \frac{d}{dt_0} P(t_0+T|t_0) =  P(t_0+T|t_0)Q(t_0+T) -Q(t_0)P_j(t_0+T|t_0) . 
   \end{equation}
To lift this computation at the level of equation \eqref{eqn:fullProbIndividual}, we compute each $P_j$ separately by numerically integrating the corresponding ODE, with initial conditions given by the identity matrix, and then take their product at each initial time step of interest, relying on the Markovian nature of the approximate single agent model.    
Note that if $t_1,\ldots,t_N$ are the fixed clock constraints, and $T_j =t_{j+1} -t_j$ the fixed interval between each clock constraints. The Kolmogorov equations are define on  the traslate clock constraints $\tilde{t}_1,,\ldots,\tilde{t}_N$ such that $\tilde{t}_i=t_0 +t_i$. In this way, we obtain $P(t_0+T|t_0)$ as a function of $t_0$. The path probability $P(s_0,t_0 \models \dta) $ can then be computed according to equation \eqref{eq:prob}. Let see the next example for more explanations.

\begin{example} 
\label{ex:st1agentFcnTime}
Consider again the running example. Fixing the time horizon $T = t_N$, and the clock constraint $\tau$, let  $t_{\tau} = t_0 + \tau$ and $T_{f} = T-\tau$. We integrate the  Kolmogorov equations (\ref{eq:forwbackKE}) for $P_{[t_0,\tau]}(t_0+\tau|t_0)$ and $P_{[\tau,T]}(t_\tau+T_{f}|t_\tau) $.
Then we have that $ P(t_0 + T|t_0)= P_{[0,\tau]}(t_0 + \tau|t_0) P_{[\tau,T]}(t_\tau+T_{f}|t_\tau)) = P_{[0,\tau]}(t_0 + \tau|t_0) P_{[\tau, T]}(t_0 + T|t_0 + \tau)  $. In Figure \ref{subfig:prop1agfcnT0}, we plot the satisfaction probability of the1gDTA property of Figure \ref{DTAprop} as function of the initial time $t_0$ for a single agent and a SIR population with $N=100$.  We can see that the satisfaction decrease for the first 5 time units and then increase until a reach a steady state around $t=50$ time units. This is in accordance with the behaviour of the population (Figure \ref{subfig:SIR}) where the number of infected rapidly increase for the first 5 time units. The property that we are verifying requires a that the agent has to be infected only after the first $10$ time units, this implies that the higher is the number of infected at time $t_0$, the lower will be the probability to satisfy the property. We can observe also that the value for $t_0=0$ is exactly the same that we obtained in the previous example computing the probability  as a function of the time horizon $T$,  fixed here to $T=300$.

\begin{figure}[!t]
\begin{center}
\subfigure[]{
\label{subfig:prop1agfcnT0}
\includegraphics[width=.47\textwidth]{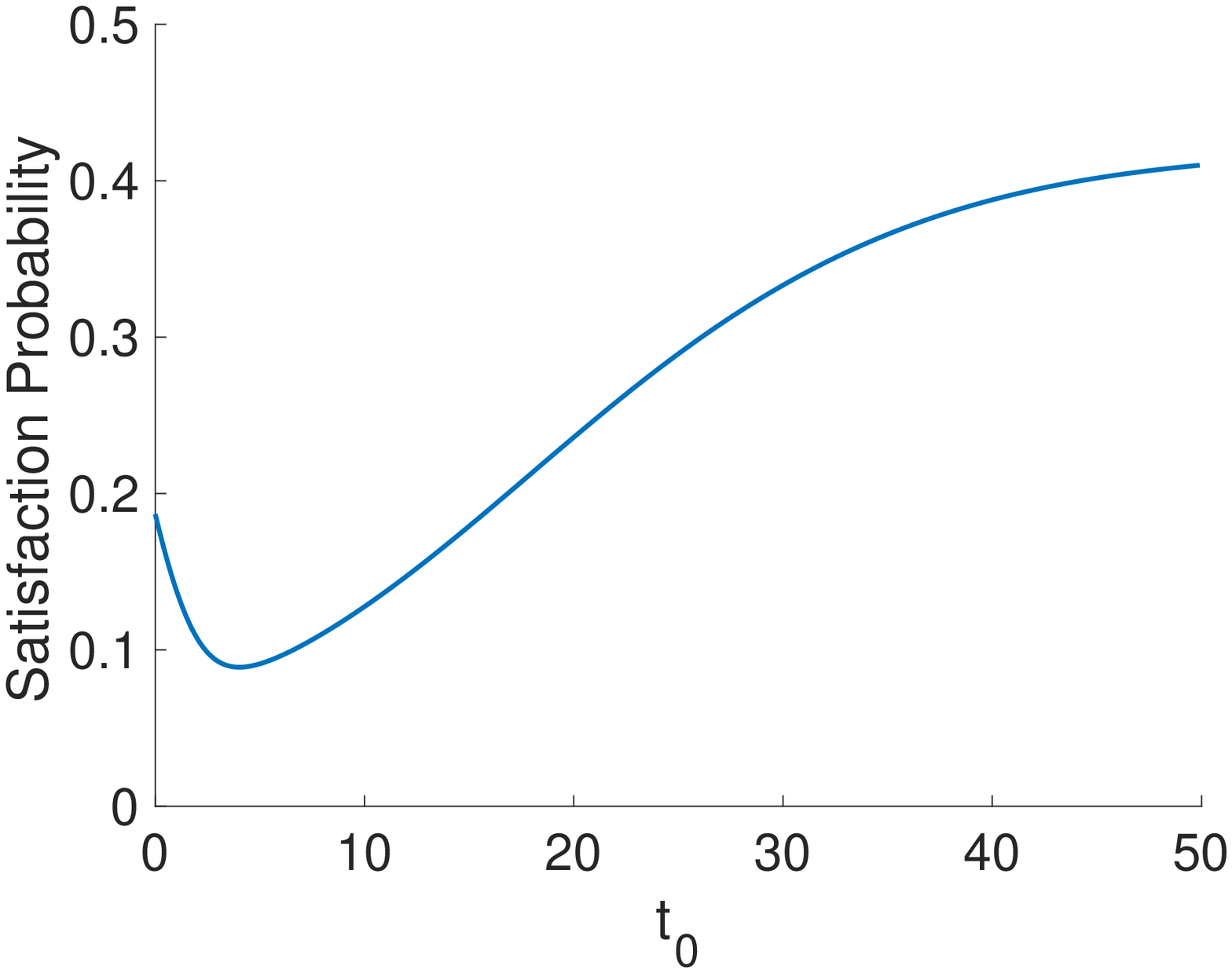}
}
 \hfill
\subfigure[]{
\label{subfig:SIR}
\includegraphics[width=.47\textwidth]{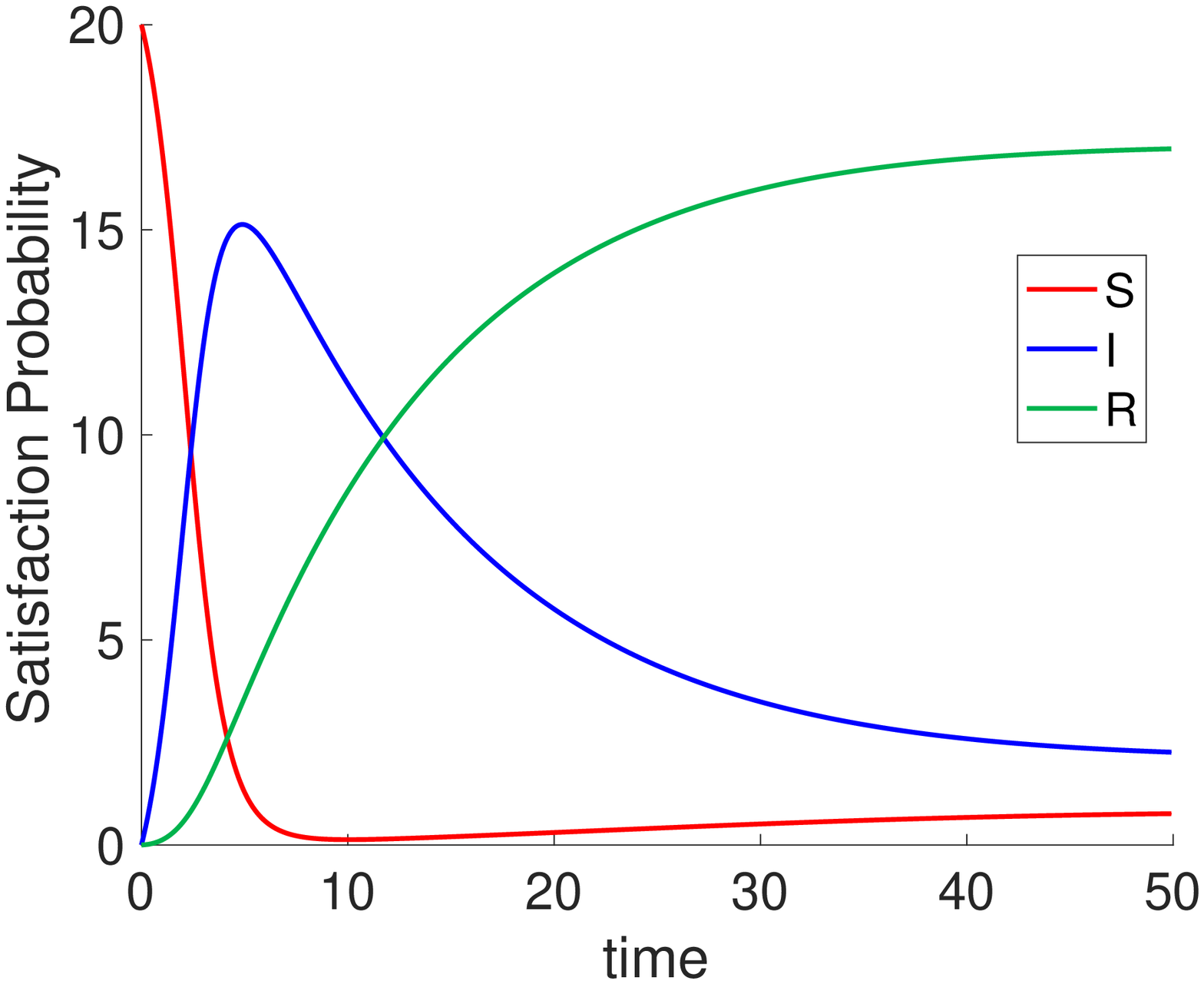}
}
\end{center}
\caption{(a) Satisfaction probability of the1gDTA property of Figure \ref{DTAprop} as function of the initial time $t_0$ for a single agent with time horizon $T=300$ and SIR population  $N=100$. The other parameters are the same as in the previous example. (b) Simulation of the SIR population model using the fluid approximation.}
\end{figure}
\end{example}

\begin{remark}
The equation \eqref{eq:forwbackKE} is a matrix valued ODE that can be solved with standard numerical routines. However, it is typically very stiff, and its direct integration may turn out to be impossible due to numerical instabilities. Typically this appears when integrating for an interval larger that a constant $T'$, which opens a way to tackle the instability using the Markov property of the process, see the appendix of \cite{bertinoro13} for a discussion of an algorithm that keeps numerical errors under control.
\end{remark}

\subsection{Model Checking 1gDTA} \label{sec:singleMC}

We now present the full model checking algorithm for the individual specification properties. The routine presented in the previous section to approximate the path probabilities is the core of the approach. In fact, the difficult property to check in the logic of Definition~\ref{def:CSLTA}, is the formula $\P^{\leq T}_{\bowtie p}\left(\dta[\Phi_1,\ldots,\Phi_k] \right)$, whose truth is easily obtained once the function 
$P(s_0,t_0 \models \dta)$ is computed by fluid approximation for the product model of the agent-property. The only extra operation that we need to solve is to check if $P(s_0,t_0 \models \dta)) \bowtie p$. This is not necessarily a trivial operation, because for nested subformulae we need to do this check for each initial time $t_0$, and there are uncountably many of them. The solution is to resort to numerical  routines that look for all the zeros of the function $P(s_0,t_0 \models \dta)-p$, possibly relying on root finding algorithms embedded in ODE solvers \cite{burden}. In this way, we can compute a boolean-valued function returning the truth value for each state and initial time $t_0$. The complete procedure 
to check $\Phi = \P^{\leq T}_{\bowtie p}\left(\dta[\Phi_1,\ldots,\Phi_k] \right)$ is sketched in Algorithm \ref{mcpath}.  It takes as input time dependent CSL-TA formulae $\Phi_1(t)$,\ldots, $\Phi_k(t)$, it performs first the structural resolution of timed-varying properties $\Phi_j(t)$, as discussed in remark \ref{rem:timeDependentDTA}, then computes the path probability $P(s_0,t_0 \models \dta)$ according to the previous section, and finally solves $P(s_0,t_0 \models \dta) \bowtie p$ to return the time dependent truth value $\Phi(t_0)$.

Algorithm \ref{mcpath} is called from the full model checking procedure, which solves the model checking problem recursively on the parse tree of a CSL-TA formula.  Boolean operations on time-dependent truth profiles $\Phi_i(t)$ are performed pointwise in time. To this end, we can rely on the algorithms for boolean signals developed in \cite{maler2004} for the logic STL. 

\begin{algorithm}
    \caption{Model checking algorithm for $\P^{\leq T}_{\bowtie p}\left(\dta[\Phi_1,\ldots,\Phi_k] \right)$}
    \label{mcpath}
    \begin{algorithmic}[1] 
        \Procedure{check}{$\P^{\leq T}_{\bowtie p}\left(\dta[\Phi_1,\ldots,\Phi_k] \right)$,$\Phi_1(t)$,\ldots,$\Phi_k(t)$,$\class$,$\pop\upsize$} 
            \State construct the structural reduction $\dta'$ of $\dta[\Phi_1,\ldots,\Phi_k]$ for the timed properties $\Phi_1(t)$,\ldots,$\Phi_k(t)$.
            \State Construct the product $\apro$ between the agent $\class$ and the property $\dta'$.  
            \State Compute the solution of mean field equations $\x(t)$
            \State Compute the solution of the Kolmogorov Equations $P(t_0+T|t_0) $ for $\apro$ and $P(s_0,t_0 \models \dta)$.
            \State Compute $\Phi(t_0) \equiv  P(s_0,t_0 \models \dta) - p \bowtie 0$
            \State \textbf{return} $\Phi(t_0)$
        \EndProcedure
    \end{algorithmic}
\end{algorithm}

\subsection{Computability and convergence} \label{sec:singleConvergence}
In this section, we briefly discuss the computability and convergence of the model checking algorithm. Computability is not straightforward, as the model of the single agent we are checking depends on the solution of the fluid or of a moment closure equation, hence standard results about CTMCs \cite{aziz} do not hold. Even more complicated is the fact that we need to compare  $P(s_0,t_0 \models \dta)$ with the threshold $p$ not for a single time point, but for uncountably many. Hence, we need conditions guaranteeing that the solution of $P(s_0,t_0 \models \dta) - p = 0$ is computable and that the number of zeros is finite.  The problem is analogous to the one discussed in  \cite{fluidmc}, hence the same recipe can be applied here. The idea is to restrict the admissible rate functions of the population model to (piecewise) real analytic functions \cite{realanalytic},\footnote{A function is  real analytic function if it admits a power series expansion.} which guarantees that the solution of the fluid and of  moment closure equations is still (piecewise) real analytic, and so are the solutions of the Kolmogorov equations for the individual agent. This in turn implies that   $P(s_0,t_0 \models \dta) - p = 0$ has a finite number of zeros, and that these zeros are computable for almost all values of the threshold $p$. Indeed, computation of the zero of a function is possible only for \emph{transversal zeros}, which are points for which the function changes sign while crossing the zero axis. This leaves out tangential zeros (the function touches zero at a minimum or maximum point). In \cite{fluidmc} it was proved that the function $P(s_0,t_0 \models \dta)-p$ has tangential zeros only for a  set of values of $p$ of Lebesgue measure zero in $[0,1]$. This justifies the introduction of the notion of \emph{quasi-computability} for the model checking problem, requiring the model checking algorithm to terminate for all but a subset of measure zero of the threshold values $p$ in the probability quantifiers.  Invoking the results of \cite{fluidmc}, we can then conclude that: 
\begin{theorem} \label{th:quasicomp}
The model checking problem based on fast simulation for local CSL-TA properties is quasi-computable, for population models with (piecewise) real analytic rate functions. \qed
\end{theorem}

An orthogonal but related issue is that of the convergence of the so computed path probabilities to the true values, i.e. of the accuracy of the approximation. In this case, we can rely on the fast simulation result (Theorem \ref{th:fastsim}), and applying similar arguments as in  \cite{fluidmc}, the following result holds true:
\begin{theorem} \label{th:convergenceIndividual}
If $\hY,s_0,t_0 \models \Phi$ is computable, then
$\hY,s_0,t_0 \models \Phi$ if and only if there is $N_0$ such that for all $N\geq N_0$,  $Y\N,s_0,t_0 \models \Phi$. \qed
\end{theorem}
The previous theorem holds for almost every formula: if $\Phi$ contains $k$  probabilistic quantifiers, we  need to discard a set of thresholds $p$ of measure zero in $[0,1]^k$.

\begin{remark}
The method presented in this section is a extension of the approach of \cite{fluidmc} for CSL to  CSL-TA. There is, however, a remarkable difference between the two approaches: nesting probabilistic operators in CSL introduces discontinuities in the function $P(s_0,t_0 \models \dta)$, which makes the verification of nested properties very challenging. Discontinuities insurge because when a subformula of the until operator changes truth value in a given state  at a given time, this induces a change of the goal or unsafe set in the corresponding reachability problem \cite{fluidmc,ctmcmc}.  CSL-TA has not such a problem: when a subformula in $\P^{\leq T}_{\bowtie p}\left(\dta[\Phi_1,\ldots,\Phi_k] \right)$ changes truth value, then there is a change in the \emph{edges} of the 1gDTA,  inducing a change in the transitions that the individual
agent synchronised with the 1gDTA can make. This changes the dynamics, i.e. the vector field (introducing a discontinuity in the derivatives), but not the value of the path probability. 
\end{remark}

\subsection{Higher-order corrections} \label{sec:singleHO}
One way to increase the accuracy of the proposed  model checking algorithm based on fluid approximation and fast  simulation is to improve the approximation of the mean $\bbE_{\X\N(t)}[Q_j(\X\N(t))] \approx Q_j(\x(t))$. In particular, we can rely on higher order corrections of the $\bbE_{\X\N(t)}[Q_j(\X\N(t))] $ using either system size expansion or moment closure techniques. 

Doing this algorithmically is straightforward: one can derive equations for the different terms in $\bbE_{\X\N(t)}[Q_j(\X\N(t))] $, which are typically monomials on the variables of  $\X\N(t)$, i.e. they correspond to mean or higher order moments (in case of non polynomial terms, a Taylor expansion is required). Then, these solutions determine corrected time dependent rates for the infinitesimal generator of the individual agent, which can be used to solve the Kolmogorov equations as for fast simulation.

In the following, we discuss how these higher order corrections work on the epidemic example,  how much they improve the accuracy, and at which computational cost.

\begin{example} 
In this example, considering the same model and property of the last Example \ref{ex:Q1agent}, we want to compare the result of the fast simulation (FS) with high order corrections. In particular, for the Moment Closure, we have considered a low dispersion of order 4 (hence we have set to zero all the moments of order greater than or equal to 5), instead for the system size expansion we used the {\it Effective Mesoscopic Rate Equation} (EMRE).
In Figure \ref{1AgfluidHighorder20}, we report the results for $N=20$ and $N=50$. As we can immediately see,  both EMRE and MM improve the estimate. In Table \ref{table:error1agHighorder}, we report the maximum and mean absolute and relative errors obtained by the FS and the higher-order approximation for $N=20,50,100$. 
We would like to remark that  the high value of the maximum relative error (RE) is misleading. In fact as it reaches such a  value at the beginning of the simulation time, when the true satisfaction probability is very close to zero, and the statistical estimate is unreliable. Note that the RE then decays very fast, as it can be seen in Figure \ref{error1ag}, where we can also clearly see that the EMRE and the MM improve sensibly the approximation. Note also how EMRE and MM are mostly effective for small populations, bringing less significant contributions for larger ones.
\label{ex:st1agenthighorder}

\begin{figure}[H]
\begin{center}
\hfill
\includegraphics[width=.49\textwidth]{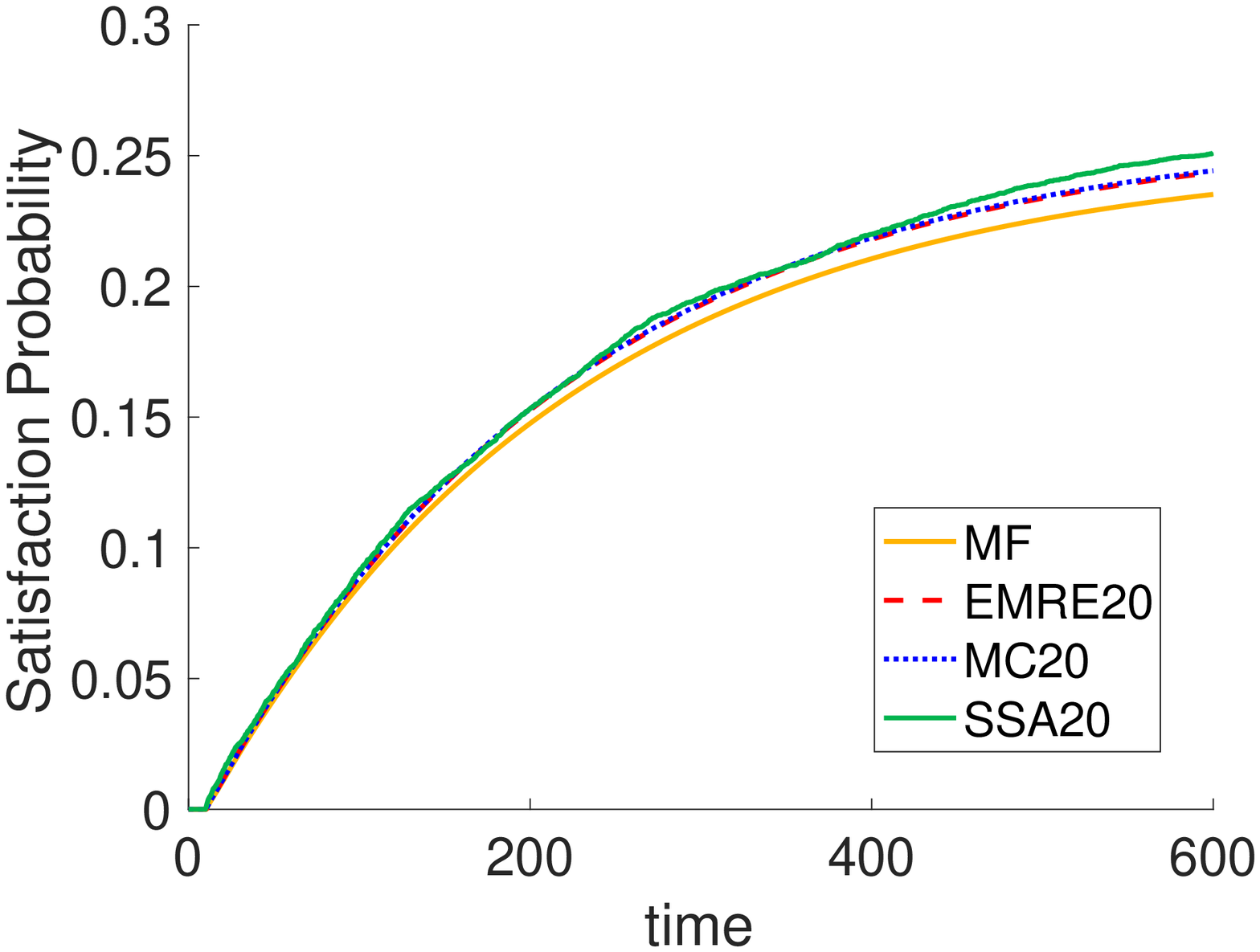}
\includegraphics[width=.49\textwidth]{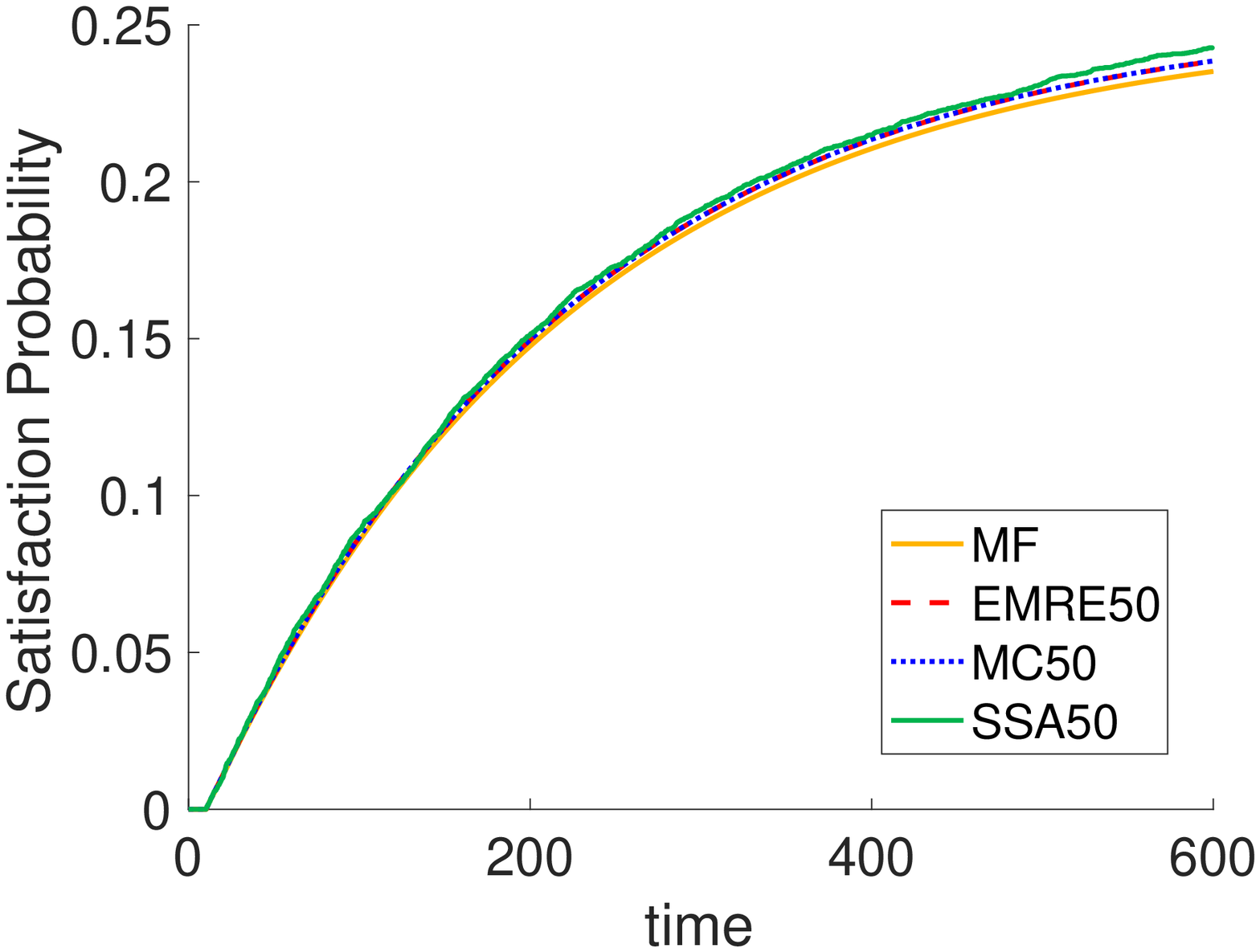}
\caption{Comparison of the results obtained by the fast simulation  (FS), the Moment Closure (MM), the Effective Mesoscopic Rate Equationa (EMRE) and the statistical estimate (SSA) of the path probabilities of the 1gDTA property of Figure \ref{DTAprop} (a), $N=20$ (left) and $N=50$(right). Parameters of the model are $\kappa_{inf}=1$, $\kappa_{patch_1}=0.08$, $\kappa_{loss}=0.01$, $\kappa_{ext}=0.1$, $\kappa_{patch_0}=0.001$.}
\label{1AgfluidHighorder20}
\end{center}
\end{figure}

\begin{table}[!t]
\begin{center}
\hfill
\begin{footnotesize}
\begin{tabular}{|c|c|c|c|c|c|c|}
\hline
    $N$   & max(erFS) & $\bbE$[erFS] & max(erEMRE) & $\bbE$[erEMRE]& max(erMM) & $\bbE$[erMM] \\ 
    \hline
    20    & 0.0159    & 0.0086        &  0.0089     &   0.0040  &  0.0088    &   0.0039 \\
    \hline
    50   &  0.0121     & 0.0062        &  0.0076     &  0.0029  &  0.0069     &  0.0025 \\
    \hline
    100   &  0.0045     &  0.0017        &  0.0056     &  0.0027&  0.0056     &  0.0028 \\
    \hline
    \hline
    $N$   & max(RerFS) & $\bbE$[RerFS] & max(RerEMRE) & $\bbE$[RerEMRE]& max(RerMM) & $\bbE$[RerMM] \\ 
    \hline
    20    &  0.8966   & 0.0584       &  0.8859     &   0.0278  &  0.8837    &   0.0274 \\
    \hline
    50   &  0.8506     & 0.0406        &  0.8447     &  0.0251  &  0.8443     &  0.0228 \\
    \hline
    100   &  0.5267     &  0.0166        &  0.5173     &  0.0227&  0.5170     &  0.0227\\
    \hline
\end{tabular}
\end{footnotesize}
\end{center}
\caption{Maximum and mean absolute and relative error on the reachability probability estimations obtained by the Fast Simulation (FS), the EMRE and the Moment Closure (MM) in the experiments of Figure \ref{1AgfluidHighorder20}.}
\vspace{-2ex}
\label{table:error1agHighorder}
\end{table}

\begin{figure}[H]
\begin{center}
\hfill
\includegraphics[width=.49\textwidth]{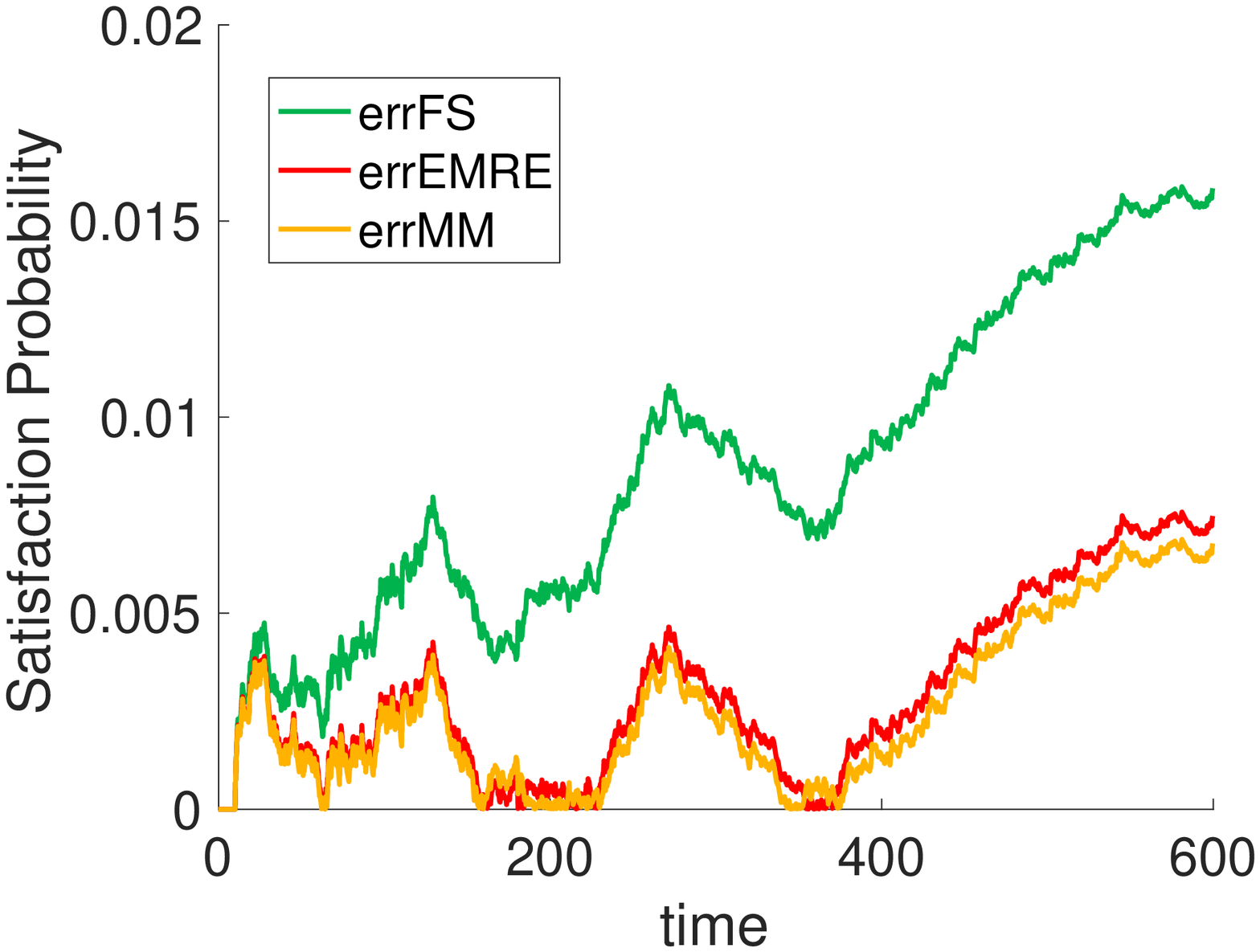}
\includegraphics[width=.49\textwidth]{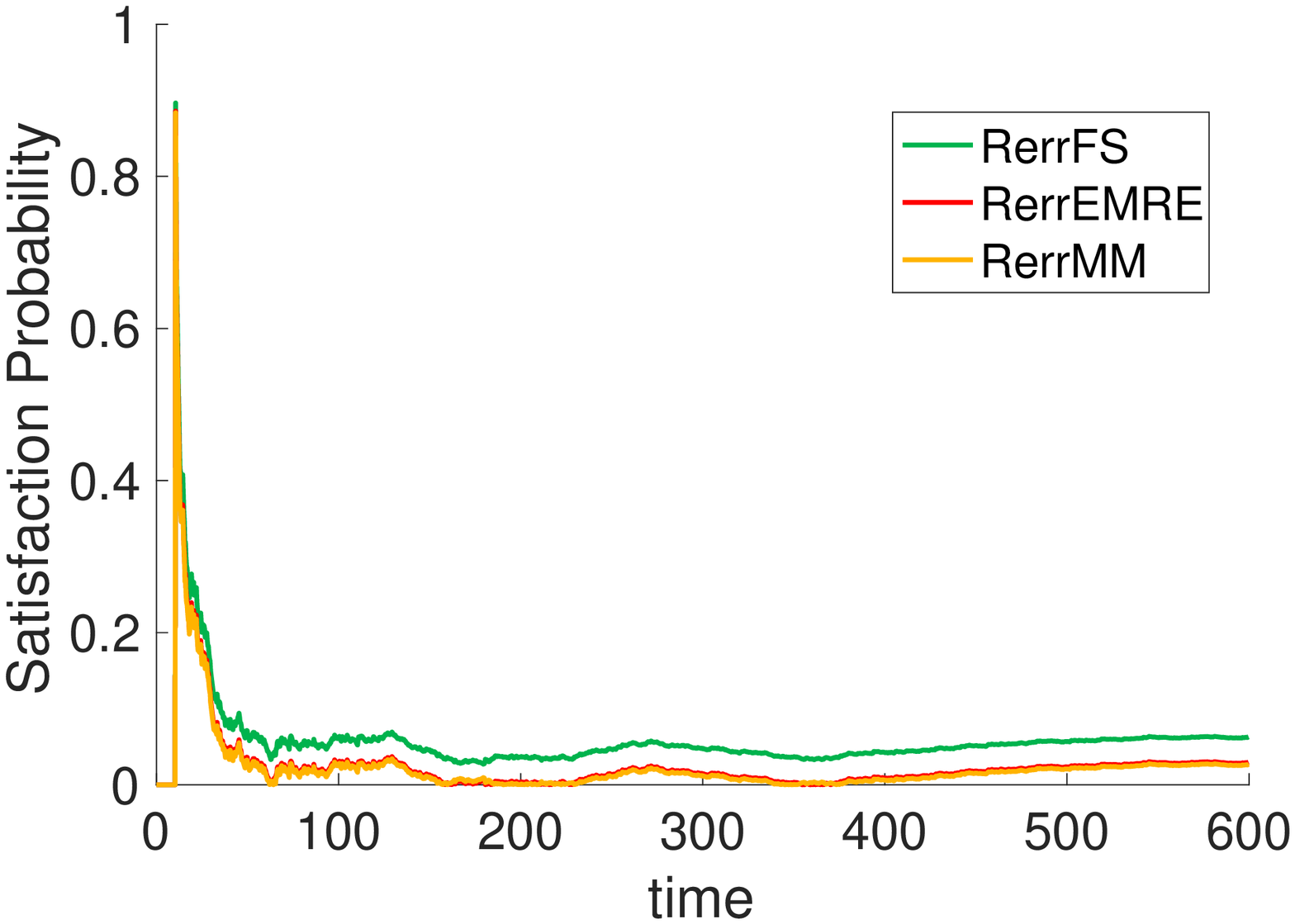}
\caption{Plot of the  absolute errors (left) and the relative errors (right) for $N=20$.}
\label{error1ag}
\end{center}
\end{figure}

\end{example}

%

\section{Model Checking Collective Properties}
\label{sec:modelchecking}

In this section, we show how to deal with the collective properties of Definition \ref{def:globalProp}. The mechanism is similar to the one for individual properties: starting from the synchronization of an agent class with a property, we can construct a new collective model, in which population variables will count how many agents of that class are in a given agent-property product state.  Once this collective model is constructed, we can use it to compute the probabilities of collective path formulae, which are the main challenge also in this case. For this, we need to rely on the linear noise approximation, or on some higher-order moment closure technique combined  with distribution reconstruction routines, like maximum entropy \cite{andreychenko2015}.
Finally, we will show how to check the other collective properties and in particular the collective state properties (boolean combinations are trivial). 
While presenting the method, we will comment also on its asymptotic correctness.

\subsection{Collective Synchronisation of Agents and Path Properties}
\label{sec:collectivesynch}

%


In order to model check collective path properties, we need to update the population model $\pop\upsize = (\class, \ptrsp\upsize, \pinst\upsize)$ so that we can count how many agents of class  $\class = (\cstsp, \ctrsp)$ satisfy a local specification  $\dta = (\lbset, \atprop, \gstsp, \ginst, \gfinst, \gtr)$. 
We do this by defining the population model associated with the local property $\gdta$ as a sequence $\gpop\upsize = (\pop_{I_{1}}\upsize, \ldots, \pop_{I_{k}}\upsize)$ of population models. Since the agent states are synchronized with the property automaton,  each transition in the population model needs to be replicated many times in the extended collective model  to account for all possible combinations of the extended local state space. Furthermore, we also need to take care of rate functions in order not to change the global rate. Recall from Section \ref{sec:singleSynch} the definition of the agent class $\apro = (\aca{1}, \ldots,\\ \aca{k+1})$ associated with the property $\gdta$,  which contains a sequence of deterministic automata $\aca{j} = (\acstsp, \actrsp_{j})$, $j = 1, \ldots, k+1$, one for each DTA with time constraint resolved. 

Let us fix the attention on the $j$-th element $\aca{j} $ in the agent class $\apro$ associated with the property $\gdta$. The state space of each $\aca{j} $ is $S\times Q$, hence to construct the global model we need  $nm$ counting variables  ($n=|S|$, $m=|Q|$), where $X_{s,q}$ counts how many agents are in the local state $(s,q)$. 
Let $\tau = (\syncset{\tau},f\upsize) \in \ptrsp\upsize$ be a global transition, apply the relabeling of action labels, according to step 1 of Section \ref{sec:singleSynch}, and focus on the synchronisation set $\syncset{\tau} = \{ s_1\xrightarrow{\alpha_{s_1}} s_1',\ldots, s_k\xrightarrow{\alpha_{s_k}} s_k'  \}$. 
We need to consider all possible ways of associating states of $Q$ with the different states $s_1,\ldots,s_k$ in $\syncset{\tau}$. Indeed, each choice $\vec{q} = (q_1,\ldots,q_k)\in Q^k$ generates a different transition in $\pop_{I_{j}}\upsize$, with synchronization set $\syncset{\tau,\vec{q}} = \{ (s_1,q_1)\xrightarrow{\alpha_{s_1}} (s_1',q_1'),\ldots, (s_k,q_k)\xrightarrow{\alpha_{s_k}} (s_k',q_k')  \}$, where $q_i'$ is the unique state of $Q$ such that $q_i\xrightarrow{\alpha_{s_i}} q_i'$. The rate function $f_{\vec{q}}\upsize$ associated with this instance of $\tau$ needs to be  a fraction of the total rate function $f\upsize$ of $\tau$, proportional, for each $i=1,\ldots,k$, to the fraction of the agents in state $s_i$ that is in the combined state $(s_i,q_i)$, accounting for the correct multiplicity in the synchronisation set. In particular, let $\kappa_{s_i}^{q_i}$ be the multiplicity with which $(s_i,q_i)$ appears in the left hand side of a rule in $\syncset{\tau,\vec{q}}$, and ${\kappa}_{s_i}$ be the multiplicity of $s_i$ as a left hand side in  $\syncset{\tau}$.  

The simplest way to proceed is to fix an ordering of the elements of $\syncset{\tau}$ and count how many ordered tuples of agents we can form in the current state $\X$ of the system, where element $j$ of the tuple is an element in state $s_j$, as specified in $\syncset{\tau}$. By doing the same for $\syncset{\tau,\vec{q}}$, and taking the ratio of the two quantities, we obtain the following formula for the rate: 
%
%
%
\begin{equation}
\label{eqn:globalRate}
f_{\vec{q}}\upsize (\pstvec) = \frac{\prod_{(s,q)\in LHS(\syncset{\tau,\vec{q}})} \frac{X_{s,q}!}{(X_{s,q}-\kappa_{s}^{q} )!}}{\prod_{s\in LHS(\syncset{\tau})} \frac{X_{s}!}{(X_{s}-\kappa_{s} )!}  } f\upsize (\pstvec) 
%
\end{equation}
where $LHS(\syncset{\tau})$ is the set containing all the states appearing on the left hand side of a rule in $\syncset{\tau}$, and similarly for $LHS(\syncset{\tau,\vec{q}})$.
Moreover, $\widetilde{\pstvec} = (X_1, \ldots, X_n)$ with $X_s = \sum_{r = 1}^{m} X_{s,r}$. Due to the restrictions enforced in Definition \ref{populationModel}, summing up the rates $f_{\vec{q}}\upsize (\pstvec)$ for all possible choices of $\vec{q}=(q_1,\ldots,q_k)\in Q^k$, we obtain $f\upsize(\widetilde{\pstvec})$:
\begin{proposition}
\label{prop:global_rates}
With the definition above, it holds that
$\sum_{\vec{q}\in Q^k} f_{\vec{q}}\upsize (\pstvec) = f\upsize (\pstvec)$, i.e.
\[ \sum_{\vec{q}\in Q^k}  \frac{\prod_{(s,q)\in LHS(\syncset{\tau,\vec{q}})} \frac{X_{s,q}!}{(X_{s,q}-\kappa_{s}^{q} )!}}{\prod_{s\in LHS(\syncset{\tau})} \frac{X_{s}!}{(X_{s}-\kappa_{s} )!}}  = 1\]
\end{proposition}

The discussion above is encapsulated into the following:
\begin{definition}[Population model associated with a local property]
\label{def:popModelProp}
The population model associated with the local property $\gdta$ is the sequence $\gpop\upsize = (\pop_{I_{1}}\upsize, \ldots, \pop_{I_{k}}\upsize)$. The elements $\pop_{I_{j}}\upsize = (\aca{j}, \ptrsp_j\upsize)$ are such that $\aca{j}$ is the $j$-th element of the agent class associated with $\gdta$ and $\ptrsp_j\upsize$ is the set of global transitions of the form $\tau_i^j = (\syncset{i}^j, f_{j,i}\upsize)$, as defined above.\footnote{Initial conditions of population models in $\gpop\upsize$ are dropped, as they are not required in the following. The initial condition at time zero is obtained from that of $\pop\upsize$ by letting $(x_0)_{s,q_0} = (x_0)_s$, where $q_0$ the initial state of $\gdta$ and $s\in \cstsp$.}
\end{definition}

\subsection{Model Checking Collective Path Properties}

Consider a population model $\pop\upsize$, for a fixed population size $N$, and a global path property $\gp{\gprop{\gdta(T)}{a}{b}}{}{\bowtie p}$. This requires us to compute the probability $\gp{\gprop{\gdta(T)}{a}{b}}{}{}$ that, at time $T$,  the  fraction of agents satisfying the local specification $\gdta$ is contained in  $[a,b]$. We will achieve this by exploiting the  construction of Section \ref{sec:collectivesynch}, according to which we obtain a sequence of population models $\gpop\upsize = (\pop_{I_{1}}\upsize, \ldots, \pop_{I_{k}}\upsize)$, synchronising local agents with the sequence of deterministic automata associated with $\gdta$. In such construction we identified  a sequence of times $0=t_0,t_1,\ldots,t_{k}=T$ and in each interval $I_j = [t_{j-1},t_j]$ the satisfaction of clock constraints does not change.

Therefore, in order to compute $\gp{\gprop{\gdta(T)}{a}{b}}{}{}$, we can rely on \emph{transient analysis} algorithms for CTMCs \cite{ctmcmc}: first we compute the probability distribution at time $t_1$ for the first population model $\pop_{I_{1}}\upsize$; then we use this result as the initial distribution for the CTMC associated with the population model $\pop_{I_{2}}\upsize$ and we compute its probability distribution at time $t_2$; and so on, until we obtain the probability distribution for $\pop_{I_{k}}\upsize$ at time $t_k = T$. 
At this point, we just need to observe that  the desired probability can be obtained by summing the probability of all those states $\pstvec \in \pstsp\upsize$ satisfying $\sum_{s\in \cstsp, q\in F} \hat{X}_{s,q} \in [a,b]$. This works because of the absorbing property of the final states in the 1gDTA (condition 2 of Definition \ref{def:1gDTA}), which guarantees that whenever an agent enters a final state of the property, it will never leave it, hence the quantity $\sum_{s\in \cstsp, q\in F} \hat{X}_{s,q} (T)$ collects the number of agents that have reached a final state of the property by time $T$.

Unfortunately, this direct, numerical approach to model checking suffers from state space explosion, which is severe even for a population size of few hundreds of individuals. 
For very large populations, when fluctuations are very small and the process behaves nearly deterministically, we could rely on fluid approximation to conclude that the probability of the path formula is approximatively equal to one if and only if  $\sum_{s\in \cstsp, q\in F} \Phi_{s,q}(T) \in (a,b)$, where $\Phi$ is the solution of the fluid equation, and to zero otherwise (excluding the border cases in which the sum equals either $a$ or $b$).
Populations of the order of few hundreds individuals, however, are too small to invoke the fluid limit in such a way, and fluctuations still play a major role. Hence, in order to apply stochastic approximation to estimate the satisfaction probability, we need to rely on a technique giving information about the distribution of the process at a given time. It is here that the Central Limit Approximation enters the picture.

The idea is simply to compute the average and covariance matrix of the approximating Gaussian Process by solving the ODEs shown in Section \ref{sec:cla}. In doing this, we have to take proper care of the different population models associated with the time intervals $I_j$. Then, we integrate the Gaussian density of the approximating distribution at time $T$ to estimate of the probability $\gp{\gprop{\gdta(T)}{a}{b}}{}{}$. The justification of this approach is in Theorem \ref{th:central}, which guarantees that the estimated probability is asymptotically correct, but in practice, we can obtain good approximations also for relatively small populations, in the order of hundreds of individuals. 

\subsubsection*{Verification algorithm by Central Limit Approximation.}
The \textit{input} of the verification algorithm is:
\begin{itemize}
\item[$\bullet$] an agent class $\class = (S, E)$ and a population model $\pop\upsize = (\class, \ptrsp\upsize, \pinst\upsize)$;
\item[$\bullet$] a local property specified by a 1gDTA $\gdta = (\lbset, \atprop, \gstsp, \ginst, \gfinst, \gtr)$;
\item[$\bullet$] a global property $\gp{\gprop{\gdta(T)}{a}{b}}{}{\bowtie p}$ with time horizon $T > 0$.
\end{itemize} 
The \textit{steps} of the algorithm are:

\begin{enumerate}

\item \textbf{Construction of the population model associated with $\gdta$}. Construct the \emph{normalised} population model $\popn\upsize = (\popn_{I_{1}}\upsize, \ldots, \popn_{I_{k}}\upsize)$ associated with $\gdta$ according to the recipe of Section \ref{sec:collectivesynch}. Then modify it by adding to its vector of counting variables $\hat{\pstvec}\upsize$ a new variable $\hat{X}_{Final}$ that keeps track of the fraction of agents entering any of the final states $(s,q)$, $q\in F$.\footnote{Namely, this variable is increased by one for each transition entering a final state, and never decreased.}

\item \textbf{Integration of the central limit equations}. 
For each $j=1,\ldots,k$,   generate and solve numerically the system of ODEs that describes the fluid limit $\fluid_j(t)$  and the Gaussian covariance $\cov_j[\linear(t)]$ for the population model $\pop_{I_{j}}\upsize$ in the interval $I_j = [t_{j-1},t_j]$, with  initial conditions $\fluid_j(t_{j-1}) = \fluid_{j-1}(t_{j-1})$ and $\cov_j[\linear(t_{j-1})]= \cov_{j-1}[\linear(t_{j-1})]$ for $j>1$, and $\fluid_1(0) = \x_0$, $\cov_1[\linear(0)] = 0$.\\
Define the population mean as $\mean\upsize[\pstvec(t)] = \size \fluid_j(t)$ and the population covariance as $\cov\upsize[\pstvec(t)] = \size \cov_j[\linear(t)]$, for $t\in I_j$. Finally, identify the component $E_{Final}\upsize[\pstvec(t)]$ and the diagonal entry $C_{Final}\upsize[\pstvec(t)]$ corresponding to $X_{Final}$.

\item \textbf{Computation of the probability}. 
Let $g(x~|~\mu,\sigma^2)$ be the probability density of a Gaussian distribution with mean $\mu$ and variance $\sigma^2$. Then, approximate $\gp{\gprop{\gdta(T)}{a}{b}}{}{}$ by 
\[ \tilde{P}\upsize_\gdta(T) = \int_{\size a}^{\size b} g(x~|~E_{Final}\upsize[\pstvec(t)],C_{Final}\upsize[\pstvec(t)])\text{d}x,\]
and compare the result with the probability bound $\bowtie p$.
\end{enumerate}
The asymptotic correctness of this procedure is captured in the next theorem, whose proof is obtained by an application of Theorem \ref{th:central}, and reported in the appendix. We denote by ${P}\upsize_\gdta(T)$ the exact value of $\gp{\gprop{\gdta(T)}{a}{b}}{}{}$   and by $\tilde{P}\upsize_\gdta(T)$ the approximate value computed by the Central Limit Approximation.

\begin{theorem}
\label{th:convergence}
Under the hypothesis of Theorem \ref{th:central}, it holds that \linebreak $\lim_{N\rightarrow\infty}\|{P}\upsize_\gdta(T) -  \tilde{P}\upsize_\gdta(T)\| = 0 $. \qed
\end{theorem}

\begin{remark}
\label{rem:FinalVar}
The introduction of the counting variable $X_{Final}$ is needed to correctly capture the variance in entering one of the final states of the property. Indeed, it holds that $X_{Final} = \sum_{s\in \cstsp,q\in F} X_{s,q}$, and in principle we could have applied the CLA to the model without $X_{Final}$, using the fact that the sum of Gaussian variables is Gaussian (with mean and variance given by the sum of means and variances of the addends). In doing this, though, we would have overestimated the variance of $X_{Final}$, because we would  implicitly take into account the dynamics within the final components, adding their variance. The introduction of $X_{Final}$, instead, avoids this problem, as its variance depends only on the events that allow the agents to enter one of the final states. 
\end{remark}

\begin{example}
\label{ex:CLA}
We  discuss now the quality of the Central Limit Approximation for mesoscopic populations from an experimental perspective. We present a detailed investigation of the behaviour of the network epidemics described in Figure \ref{SIRagent}.
We consider the two local properties expressed as 1gDTAs shown in Figure \ref{DTApropExa}. The first property $\gdta_1$ has no clock constraints on the edges of the automaton, therefore the 1gDTA reduces to a DFA. The property is satisfied  if an infected node is  patched before being able to infect other nodes in the network,  thus checking the effectiveness of the antivirus deployment strategy. The second property $\gdta_2$, instead, is properly timed, and it is the same property that we use in the previous section for the single agent. 
It is satisfied  when a susceptible node is infected by an internal infection after the first $\tau$ units of time. The corresponding global properties that we consider are $\gp{\gpropg{\gdta_1(T)}{\alpha_1}}{}{\bowtie p}$ and $\gp{\gpropg{\gdta_2(T)}{\alpha_2}}{}{\bowtie p}$, where $\alpha_i$ is the fraction of agents that has to satisfy $\gdta_i$. 

In Fig.~\ref{fig:glob1} we report the final step of the synchronisation procedure for the first property $\gdta_1$. The synchronisation procedure of $\gdta_2$ was already reported in the Example~\ref{ex:syncA_DTA}, Fig.~\ref{fig:sync}. Note that the state space of $\mathscr{P}$ is $S \times Q_1$ where $S =\{S, I, R\} $ is the state space of the SIR automata and $Q_1= \{ q_b, q_0, q_f\}$ is the state space of the $\gdta_1$ local property; hence, the global model $\mathscr{P}$ has $nm = 9$ counting variables ($n=|S|$, $m=|Q_1|$).

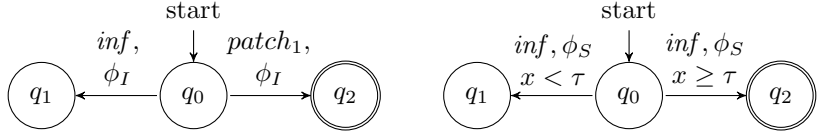
\begin{figure}[!t]
\begin{center}
\begin{tikzpicture}[on grid, shorten <=1pt, >=stealth', auto]
  \node at (2,0)   [initial above,state] (q0)   {$q_0$};
  \node at (0,0)   [state] (q1)   {$q_1$};
  \node at (4,0)   [state, accepting] (q2)   {$q_2$};  
  \path [->] (q0)   edge [ above, align=center ]  node { $\mathit{inf},$\\$\phi_I$} (q1)
		(q0)   edge [ above, align=center ]  node {$patch_1,$\\ $\phi_I$} (q2);    
\end{tikzpicture}
\ \ \ \ \ \ 
\begin{tikzpicture}[on grid, shorten <=1pt, >=stealth', auto]
  \node at (2,0)   [initial above,state] (q0)   {$q_0$};
  \node at (0,0)   [state] (q1)   {$q_1$};
  \node at (4,0)   [state, accepting] (q2)   {$q_2$};  
  \path [->] (q0)   edge [ above, align=center ]  node { $\mathit{inf},\phi_S$\\ $x < \tau $} (q1)
		(q0)   edge [ above, align=center ]  node {$\mathit{inf},\phi_S$\\ $x \geq \tau$} (q2);    
\end{tikzpicture}

\end{center}
\caption[Two 1gDTA specifications.]{The 1gDTA specifications experimentally analysed in Section \ref{sec:l2g:results}.}
\label{DTApropExa}
\end{figure}

\begin{figure}[H]
\begin{center}
$\mathscr{P}_{} =$
\begin{footnotesize} 
\begin{tikzpicture}[on grid, shorten <=1pt, >=stealth, auto]
  \node at (4,2)       [state] (s42)  {${\bm I_{0}}$};
  \node at (5,4)       [state] (s54)  {${\bm R_{0}}$};
  \node at (3,4)       [state] (s34)  {${\bm S_{0}}$};
  \node at (6,0)       [state] (s60)  {${\bm I_{b}}$};
  \node at (8,0)       [state] (s80)  {${\bm R_{b}}$};
  \node at (7,2)       [state] (s72)  {${\bm S_{b}}$};
  \node at (0,0)       [state, accepting] (s00)  {${\bm I_{f}}$};
  \node at (2,0)       [state, accepting] (s20)  {${\bm R_{f}}$};
  \node at (1,2)       [state, accepting] (s12)  {${\bm S_{f}}$};

  \path [->] (s42)  edge [bend left, left, near end]     node {$inf_{I,0}$}      (s60)
                 (s42)  edge [bend right, right, near end]       node {\textcolor{red}{$patch_{1,0}$}}      (s20)  
                 (s54)  edge [bend right, above]      node {$loss_{0}$}    (s34)
                 (s34)  edge [bend right, below, near end]      node {$patch_{0,0}$}    (s54)
                 (s34)  edge [bend right, left]      node {$inf_{S,0}$}    (s42)
                  (s34)  edge [right, right, near end]      node {$ext_{0}$}    (s42)
                 (s80)  edge [bend right, right, near end]     node {$loss_{b}$}    (s72)
                 (s72)  edge [left, very near end]     node {$patch_{0,b}$}    (s80)
                 (s72)  edge  [bend right, left, very near start]        node {$inf_{S,b}$}    (s60)
                 (s72)  edge [right]      node {$ext_{b}$}    (s60)
                 (s60) edge [bend right, below]      node {$patch_{1,b}$}    (s80)
                 (s60)  edge [loop left, below]       node {$inf_{I,b}$}   ()
                 (s20)  edge [bend right, right, near end]     node {$loss_{f}$}    (s12)
                 (s12)  edge [left, very near end]     node {$patch_{0,f}$}    (s20)
                 (s12)  edge  [bend right, left, very near start]        node {$inf_{S,f}$}    (s00)
                 (s12)  edge [right]      node {$ext_{f}$}    (s00)
                 (s00) edge [bend right, below]      node {$patch_{1,f}$}    (s20)
                 (s00)  edge [loop left, below]       node {$inf_{I,f}$}   ();
\end{tikzpicture}
\end{footnotesize}
\end{center}
\caption{  (a) A DFA specification.  (b) Synchronisation of the SIR automata described in Figure \ref{SIRagent} and the  DFA specification report in subfigure (a).}
\label{fig:glob1}
\end{figure}
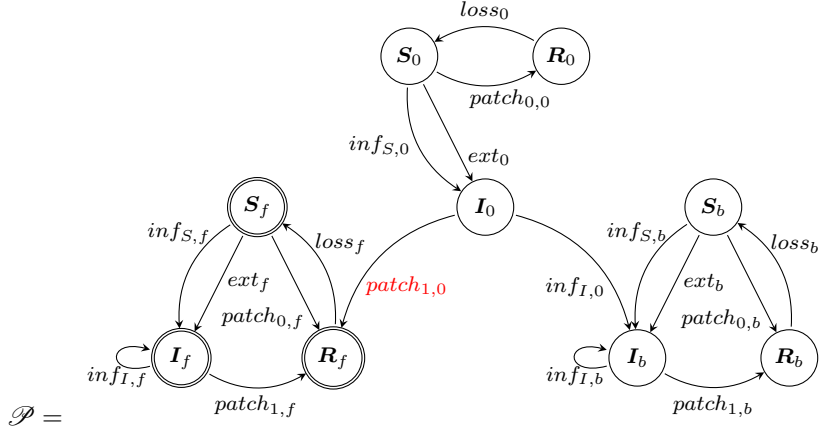

The population model associated with the local property $\gdta_1$ is then 
equal to $\gpop\upsize = ( \aca, \ptrsp\upsize)$, where $ \ptrsp\upsize$ is the set of global transitions.
We modify it adding a new variable $\hat{X}_{Final}$ that keeps track of the fraction of agents entering in the final state $(I, q_f) =I_f$; this can happen only with the transition $I_0 \xrightarrow{inf} R_f$. Hence, whenever such a transition fires, we also increase appropriately the value of $\hat{X}_{final}$, by a straightforward modification of the update vector associated with such a transition. Then, we integrate the central limit equations and we identify the component $E_{Final}\upsize[\pstvec(t)]$ and the diagonal entry $C_{Final}\upsize[\pstvec(t)]$ corresponding to $X_{Final}$. A similar procedure is done for the property $\gdta_2$.

In Figure \ref{experimentl2g}, we show the approximate probability $\tilde{P}\upsize_{\gdta_{i}}(T)$ of $\gp{\gprop{\gdta_i(T)}{\alpha_i}{1}}{}{}$)  as a function of the time horizon $T$, for different values of $N$ and a specific configuration of parameters ($\kappa_{\mathit{inf}}=0.05$, $\kappa_{patch_1}=0.02$, $\kappa_{loss}=0.01$, $\kappa_{ext}=0.05$, $\kappa_{patch_0}=0.001$, $\alpha_1=0.5$, $\alpha_2=0.2$). The CLA is  compared with a statistical estimate, obtained from 10000 simulation runs.   
As we can see, the accuracy in the transient phase increases rapidly with $N$, and  the estimate is  very good for both properties already for  $N=100$. The same parameter configuration was used to obtain the computational costs (in Seconds), showed in Table \ref{table:speedupl2g}. As we have seen, by definition the Central Limit Approximation (CLA) is independent of the population size $N$ and its computational costs is hundreds of times less than that of the statistical estimate (the Gillespie Algorithm) for both the first and the second properties.  The values shown in Figure \ref{experimentl2g}  can then be easily compared  with the probability bound $\bowtie p$, to check the satisfiability of the property $\gp{\gpropg{\gdta_i(T)}{\alpha_i}}{}{\bowtie p}$.

\begin{figure}[!t]
\begin{center}
\includegraphics[width=.49\textwidth]{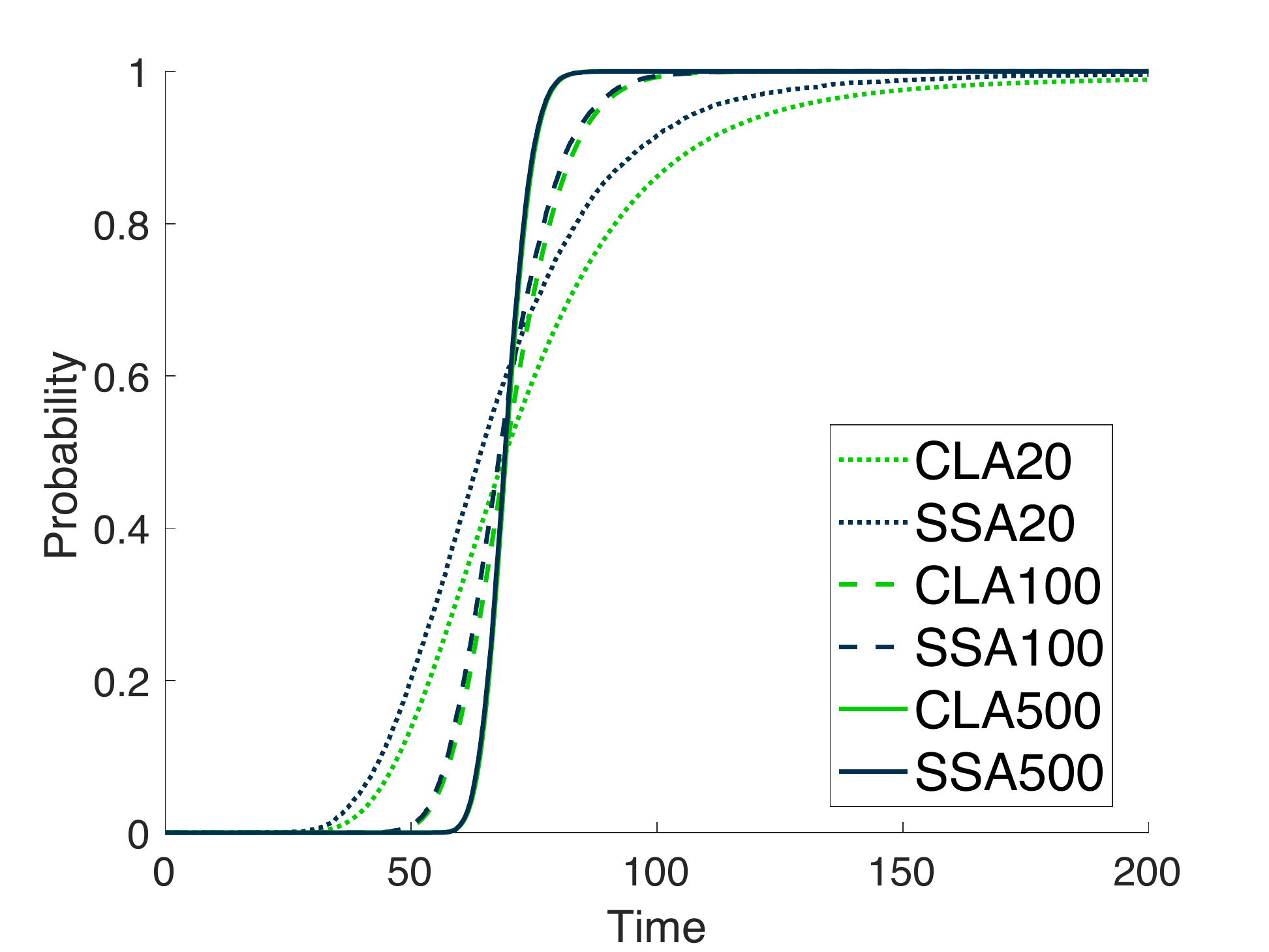}
\includegraphics[width=.49\textwidth]{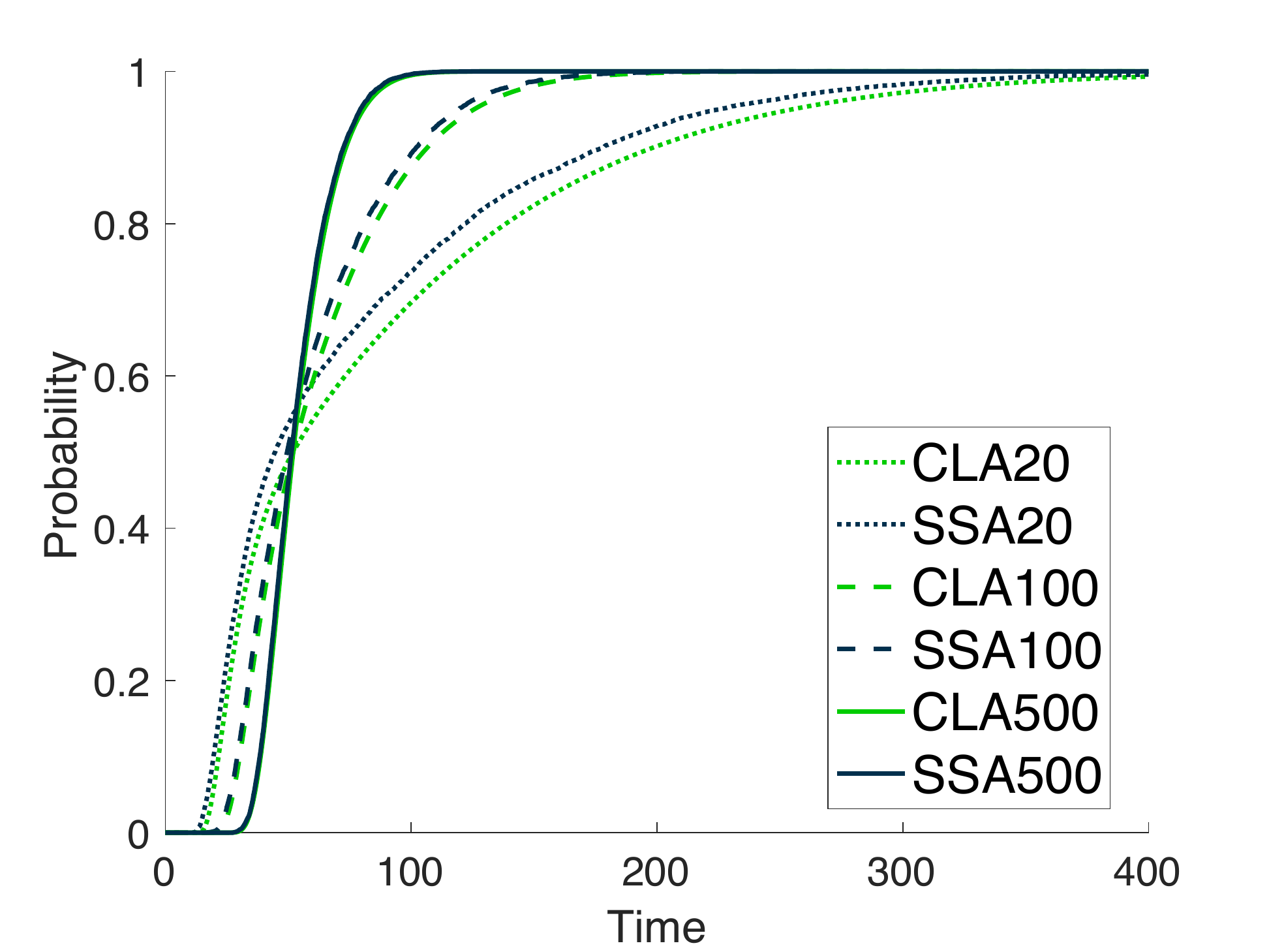}
\end{center}

\caption[Results obtained by the Central Limit Approximation and the Gillespie's Statistical Algorithm in the validation of Local-to-Global Properties.]{Comparison of Central Limit Approximation (CLA) and a statistical estimate (using the Gillespie algorithm, SSA) of the path probabilities of the 1gDTA properties of Figure \ref{DTApropExa} computed on the network epidemic model for different values of the population size $N$.}
\label{experimentl2g}
\end{figure}


\begin{table}[!t]
\begin{center}
\textbf{First Property}\\
\begin{small}
\begin{tabular}{|c||c|c|c|}
\hline
$N$ & SSAcost & CLAcost & Speedup  \\ 
\hline
    20 & 22.4114 & 0.0618 & 362.6440\\
    \hline
    50   & 23.3467  & 0.0618 & 377.7783\\
    \hline
 100   & 24.2689  & 0.0618 & 392.7006\\
 \hline
  200 &  26.1074  & 0.0618 & 442.4498\\
  \hline
  500 & 28.8754   & 0.0618 & 467.2395\\
\hline
\end{tabular}
\end{small}

\vspace{0.2cm}
\textbf{Second Property}\\
\begin{small}
\begin{tabular}{|c||c|c|c|}
\hline
$N$ & SSAcost & CLAcost & Speedup\\ 
\hline
    20 &  32.0598 & 0.3035 & 105.6336\\
    \hline
    50  & 29.0915 & 0.3035 & 95.8534\\
    \hline
 100     & 28.8651 & 0.3035 & 95.1074\\
 \hline
  200  &   33.9825 & 0.3035 & 111.9687\\
  \hline
  500   & 43.4737 & 0.3035 & 143.2412\\
\hline
\end{tabular}
\end{small}
\end{center}
\caption[Computational costs of the Central Limit Approximation in the validation of Local-to-Global Properties.]{Average computational costs (in Seconds) of the Gillespie Algorithm (SSAcost) and the Central Limit Approximation (CLAcost), and the relative SpeedUp (CLAcost/SSAcost). The data are shown as a function of the population size $N$ (by definition the CLA is independent of $N$).}
\label{table:speedupl2g}
\end{table}

Furthermore, in order to check more extensively the quality of the approximation also as a function of the system parameters, we  ran the following experiment. We considered five different values of $N$ ($N=20,50,100,200,500$). For each of these values, we randomly chose 20 different combinations of parameter values, sampling uniformly from:  $\kappa_{\mathit{inf}} \in [0.05, 5]$, $\kappa_{patch_1} \in [0.02, 2]$, $\kappa_{loss} \in [0.01, 1]$, $\kappa_{ext} \in [0.05, 5]$, $\kappa_{patch_0} \in [0.001, 0.1]$, $\alpha_1 \in [0.1, 0.95]$, $\alpha_2 \in [0.1, 0.3]$.
For each parameter set, we compared the CLA of the probability of each global property with a statistical estimate (from 5000 runs), measuring the error in a grid of 1000 equi-spaced time points. We then computed the maximum error and the average error. In Table \ref{table:errorl2g}, we report the mean  and  maximum values of these quantities over the 20 runs, for each considered value of $N$. We also report the error at the final time of the simulation, when the probability has stabilised to its limit value.\footnote{For this model, we can extend the analysis to steady state, as the fluid limit has a unique, globally attracting steady state. This is not possible in general, cf. \cite{tutorial}.} It can be seen that both the average and the maximum errors decrease with $N$, as expected, and are already quite small for $N=100$ (for the first property, the maximum difference in the path probability for all runs is of the order of 0.06, while the average error is 0.003). For $N=500$, the CLA is practically indistinguishable from the (estimated) true probability. For the second property, the errors are slightly worse, but still reasonably small.

\begin{table}[!t]
\begin{center}
\textbf{First Property}
\begin{small}
\begin{tabular}{|c||c|c|c|c|c|c|}
\hline
$N$ & MaxEr  & $\bbE$[MaxEr] & Max$\bbE$[Er] & $\bbE$[$\bbE$[Er]] & MaxEr($T$) & $\bbE$[Er($T$)] \\ 
\hline
    20  &  0.1336  &  0.0420  &  0.0491 &   0.0094  &  0.0442  &  0.0037 \\
    \hline
    50   & 0.0866   & 0.0366  &  0.0631  &  0.0067  &  0.0128   & 0.0018 \\
    \hline
 100   & 0.0611   & 0.0266   & 0.0249   & 0.0030  &  0.0307   & 0.0017 \\
 \hline
  200 & 0.0504   & 0.0191  &  0.0055   &  0.0003  &  0.0033  &  0.0002 \\
  \hline
  500 & 0.0336   & 0.0120 &   0.0024   &  0.0003 &   0.0002  &  9.5e-6 \\
\hline
\end{tabular}
\end{small}

\vspace{0.2cm}
\textbf{Second Property}
\begin{small}
\begin{tabular}{|c||c|c|c|c|c|c|}
\hline
$N$ & MaxEr  & $\bbE$[MaxEr] & Max$\bbE$[Er] & $\bbE$[$\bbE$[Er]] & MaxEr($T$) & $\bbE$[Er($T$)] \\ 
\hline
    20 &  0.2478 &    0.1173   & 0.1552 &   0.0450  &  0.1662 &   0.0448 \\
    \hline
    50  & 0.2216 &    0.0767   & 0.1233 &    0.0340 &    0.1337   & 0.0361  \\
    \hline
 100    & 0.1380   & 0.0620 &    0.0887 &    0.0216 &    0.0979 &    0.0208 \\
 \hline
  200  & 0.1365   & 0.0538 &    0.0716   & 0.0053 &    0.0779  &  0.0162 \\
  \hline
  500   & 0.1187  &  0.0398   & 0.0585   &  0.0100  &  0.0725 &    0.0108  \\
\hline
\end{tabular}
\end{small}
\end{center}
\caption[Errors obtained by the Central Limit Approximation in the validation of Local-to-Global Properties.]{Errors obtained by the Central Limit Approximation in the validation of Local-to-Global Properties. Maximum  and mean of the maximum error (MaxEr, $\bbE$[MaxEr]) for each parameter configuration;  maximum and mean of the average error with respect to time (Max$\bbE$[Er]), $\bbE$[$\bbE$[Er]])  for each parameter configuration; maximum and average error at the final time horizon $T$ (MaxEr($T$), $\bbE$[Er($T$)] ) for each parameter configuration. Data is shown as a function of the network size $N$. }
\label{table:errorl2g}
\end{table}

Finally, we considered the problem of understanding what are the most important aspects that determine the error. To this end, we regressed the observed error against the following features: estimated probability value by CLA, error in the predicted average and variance of $X_{Final}$ (between the CLA and the statistical estimates), and statistical estimates of the mean, variance, skewness and kurtosis of $X_{Final}$. We used Gaussian Process regression with Adaptive Relevance Detection (GP-ADR, \cite{gp}), which performs a regularised regression searching the best fit on an infinite dimensional subspace of continuous functions, and permitted us to identify the most relevant features by learning the hyperparameters of the kernel function. We used both a squared exponential kernel, a quadratic kernel, and a combination of the two, with a training set of 500 points, selected randomly from the experiments performed. The mean prediction error on a test set of other 500 points (independently of $N$) is around  0.015 for all the considered kernels. Furthermore, GP-ADR  selected as most relevant the quadratic kernel, and in particular the following two features: the estimated probability and the error in the mean of $X_{Final}$.  This suggests that moment closure techniques improving the prediction of the average can possibly reduce the error of the method.
\end{example}

\subsubsection{Finite-Size Threshold Correction}
Results obtained by CLA can be further improved for small values of $N$ by introducing a correction on the thresholds $a$ and $b$ of a property $\gp{\gprop{\gdta(T)}{a}{b}}{}{\bowtie p}$, taking into account the discrepancy between the discrete nature of population counts and its continuous approximation. To better understand the correction, we will illustrate on a property of the form  $\gp{\gpropg{\gdta(T)}{\alpha}}{}{\bowtie p}$. In the algorithm presented above, the CLA approximation works by integrating the Gaussian approximation of the variable $X_{final}$, as computed by CLA, from $\alpha N$ to infinity. However, for small $N$, in this way we neglect the discrete nature of the state space. Suppose we would like to compute the probability of $X_{final} = i$. Using the Gaussian approximation, we would always obtain zero, unless we integrate in a region around $i$. The obvious candidate  is $[i-\frac{1}{2},i+\frac{1}{2}]$, which correspond to a partition of the interval $[0,N]$ into subintervals of the form $[i-\frac{1}{2},i+\frac{1}{2}]$\footnote{The extremes $0$ and $N$ has to be treated in a special way: $(-\infty,\frac{1}{2}]$ for 0 and $[N-\frac{1}{2},\infty)$ for $N$}.
Following this line of reasoning, instead of integrating the Gaussian approximation for $X_{final}$ from $\alpha N$, we should start from $j-\frac{1}{2}$, where $j$ is the smallest integer greater than or equal to $\alpha N$, i.e. $j=\lceil \alpha N \rceil$. Note that $j$ is the smallest value that $X_{final}$ can take to satisfy the property, when verifying it in the discrete stochastic model. Similarly, when dealing with properties of the form $\gp{\gpropl{\gdta(T)}{\alpha}}{}{\bowtie p}$, we would need to integrate up to $\lfloor \alpha N  \rfloor + \frac{1}{2}$, combining the two corrections with dealing with threshold intervals $[a,b]$.  In several experimental tests, we observed that this simple correction improves considerably the approximation,  becoming less significant for large $N$. 

\begin{example}
\label{ex:alphacorrect}
In Figure~\ref{fig:ln_cor} we see the correction at work for $N=20$ and the first property of Example \ref{ex:CLA}, in which $\alpha = 0.5$, hence $\lceil \alpha N \rceil = 10$. We can see what happens if we integrate from 9.5 instead of 10.  Integrating from 10, some probability mass is lost, and the CLA under-approximates the true solution. The correction allows us to recover some of this lost mass, improving considerably the quality of the approximation. 

\begin{figure}[!t]
\begin{center}
\includegraphics[width=.7\textwidth]{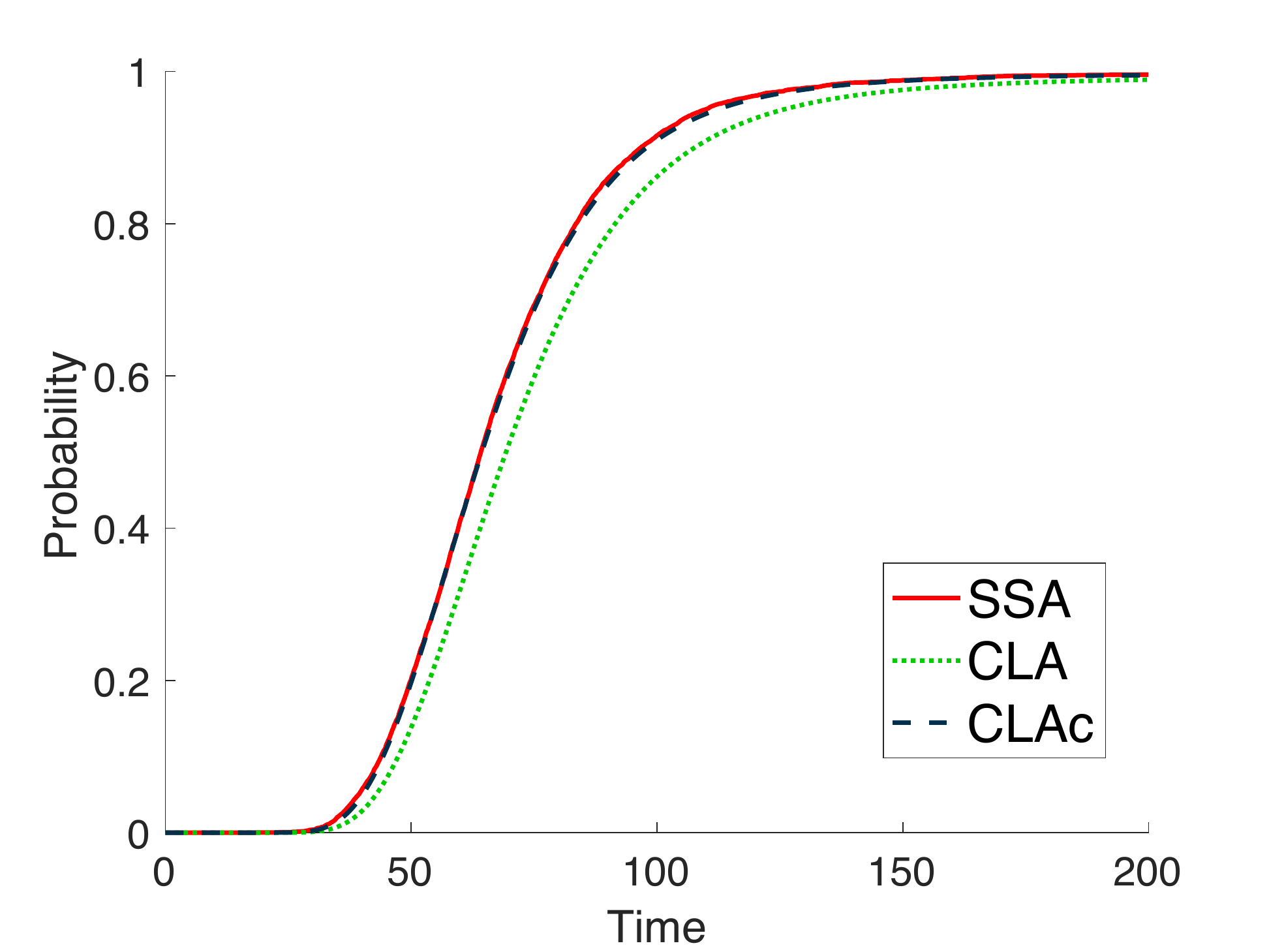}
\end{center}
\caption{Comparison of a statistical estimate (using the Gillespie algorithm, SSA, Central Limit Approximation (CLA), and the CLA with the finite-size threshold correction (CLAc) for the first property of Example \ref{ex:CLA}, $N=20$ and $\alpha = 0.5$.  }
\label{fig:ln_cor}
\end{figure}

\end{example}

\subsubsection{Verification algorithm by  Higher Order Approximations.} The method presented above relies on the central limit approximation, hence it works well when this gives an accurate description of the dynamics at time $T$. In particular, if the distribution of $X\upsize_{final}(T)$ is skewed, or deviates significantly from a Gaussian, or the fluid approximation gives a poor estimate of the mean, then this approach will give poor results. 

One way to improve the accuracy of central limit is to use higher order approximations, either higher order system size expansion or moment closure techniques. In order to use these techniques to approximately verify our property, however, we need to know how the quantity $X\upsize_{final}$ is distributed at time $T$. Unfortunately, the Central Limit Approximation is quite special in this case: in fact, it entails that the distribution of $X\upsize_{final}$ is Gaussian, hence we need just mean and covariance to characterise it.  Higher-order approximations, instead, have no closed form solution for the distribution of $X\upsize_{final}$, and they are used to access informations about its moments. Unfortunately, even the knowledge of all moments is not enough to uniquely identify a distribution. 

To tackle this problem and construct a plausible probability density function for  $X\upsize_{final}$, we need to apply some moment reconstruction technique, taking a finite number of moments of a distribution and producing a plausible approximation. In this work,
we leverage an advanced information theoretic \textit{moment-reconstruction} technique based on the \textit{maximum entropy principle} \cite{abramov2010,andreychenko2015,chapter2015}, which we exploited already in \cite{epew}, and which is of quite common usage in systems biology and in population models \cite{andreychenko2015}.
More specifically, the idea is to find the distribution $p(x)$ that maximises the entropy 
\[ p = \text{argmax}_q H[q] = -\int q(x)\log q(x) dx,  \]
subject to moment matching constraints  $\bbE_q[x^k] = \int x^k q(x)dx = \mu_k$, $k=1,\ldots, m$ where $\mu_k$ are given non-centred moments, and $\int q(x)dx  =1$, $q(x) \geq 0$.
By introducing Lagrange multipliers and applying the Kuhn-Tucker theorem \cite{berger_maximum_1996}, one can show that the solution to the distribution reconstruction problem takes the form 
\[ p(x) = \frac{1}{Z} \exp \left( -\sum\limits_{k=1}^{m} \lambda_k x^k \right), \]
where $Z$ is the partition function, i.e. the normalization constant making $p$ a distribution, and $\lambda_i$ are obtained by numerically minimising the dual formulation of the optimisation problem, namely the convex function
\[ \Psi(\lambda)=\ln Z + \sum\limits_{k=1}^{m} \lambda_k \mu_{k}. \]
More details can be found in \cite{abramov2010,andreychenko2015,chapter2015}.

Hence, to improve the estimation of  ${P}\upsize_\gdta(T)$ given in the previous section, we compute moments up to order $m$ by solving moment closure or higher order system size expansion equations, and the apply the maximum entropy moment reconstruction discussed above.   Typically, $m$ is not very large,  usually ranging between 2 and 8, with a typical value of 4 (using  moment closure equations truncated at order 5, to have a more accurate estimate of the fourth order moment). This is also due to the fact that high order moment equations tend to be stiff and difficult to integrate numerically \cite{schnoerr2016,andreychenko2015}. The drawback of this approach is that it requires the solution of a multidimensional optimization problem, and then the numerical integration of the so obtained function $p(x)$ in the range $[a,b]$ specified by the property. 
It is worth noting that the solution to maximum entropy for the first two moments only is given by a Gaussian distribution. Hence, stopping at order two gives a fast way to estimate the probability ${P}\upsize_\gdta(T)$, by using corrected mean and variance with respect to those of linear noise. If we use the system size expansion correction, we still retain convergence as the population size diverges. The behaviour of other moment closures techniques for large populations, instead, is less clear \cite{schnoerr2014}.

\begin{example}
\label{ex:highorderPop}
Let us consider again the untimed property of  Example~\ref{ex:CLA} with the same parameters, reported in the caption of the figure. For moment closure, we have considered a low dispersion of order 4, hence we have set to zero all the moments of order greater or equal to 5. 
In Figure \ref{fig:highorderMC_IOS}, we compare the results obtained for the CLA and the Gillespie's statistical estimates (with 10000 runs) (SSA), with the probabilities estimated by the IOS and the MC, for  $N=20$, without (left) and with (right) finite-size correction of the thresholds. 
In this setting, the performance of the three types of approximation (CLA, IOS and MC) is comparable. IOS and MC show a little improvement over CLA. This is better seen in Figure \ref{fig:highorderMC_IOS} (left), where we did not use the finite-size correction of the threshold, hence curves are more separated. 
\\
In Figure \ref{fig:ios}, instead, we  compare the results obtained for the CLA and the Gillespie's statistical estimates (with 10000 runs) (SSA), with the probabilities estimated by  EMRE and  IOS, for a different parameter set,  population size $N=20$, and finite-size correction of the threshold. We can see that in this case the CLA is not very accurate, while EMRE and IOS improve considerably the estimate. In this figure we have not reported the results of MC with maximum entropy reconstruction, due to numerical instabilities in the optimization phase. This is a known issue with the maximum entropy method , and a more careful implementation of the optimization is needed to circumvent such effects, a task which is beyond the scope of this paper.

\begin{figure}[t]
\begin{center}
\includegraphics[width=.49\textwidth]{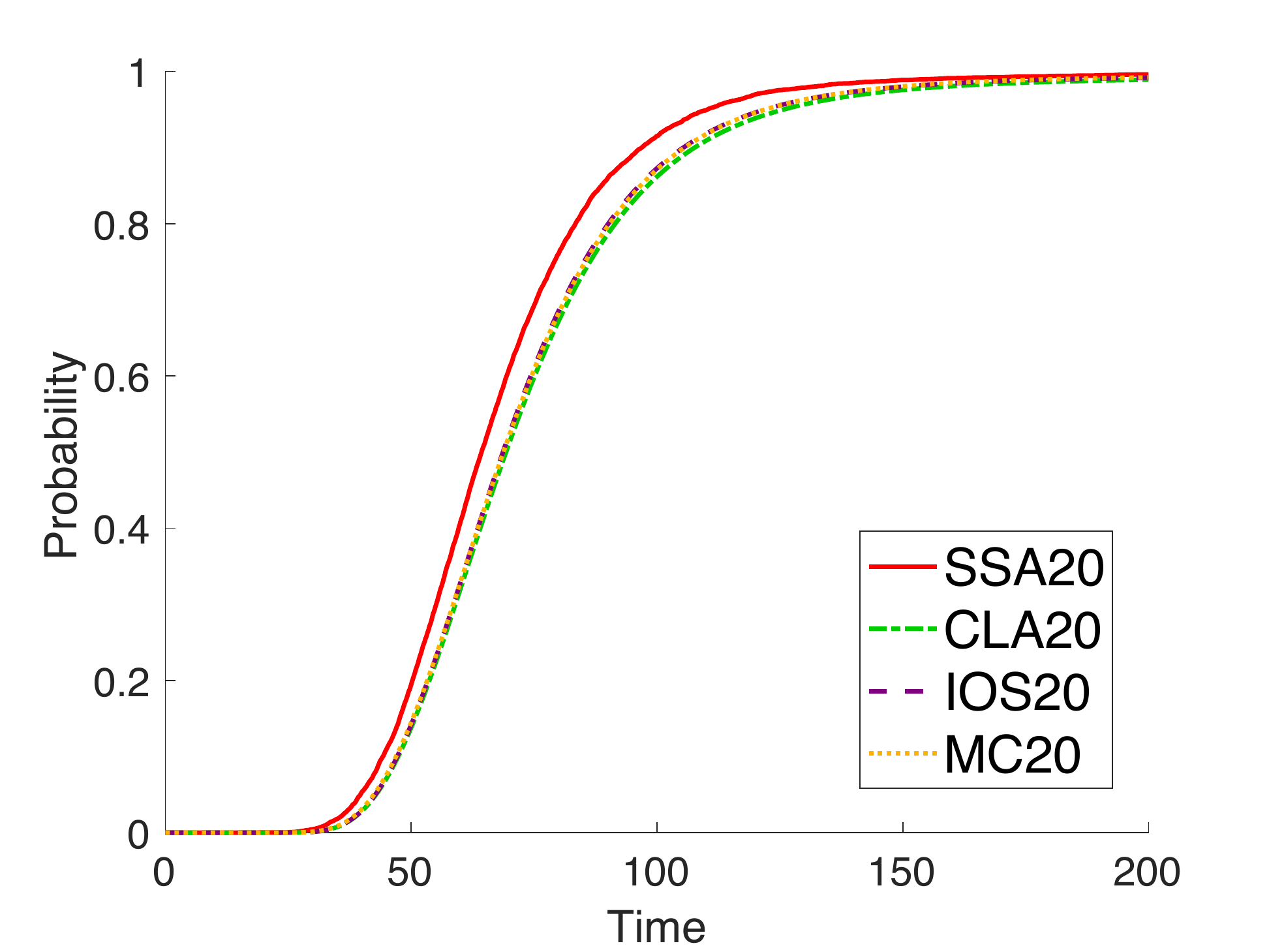}
\includegraphics[width=.49\textwidth]{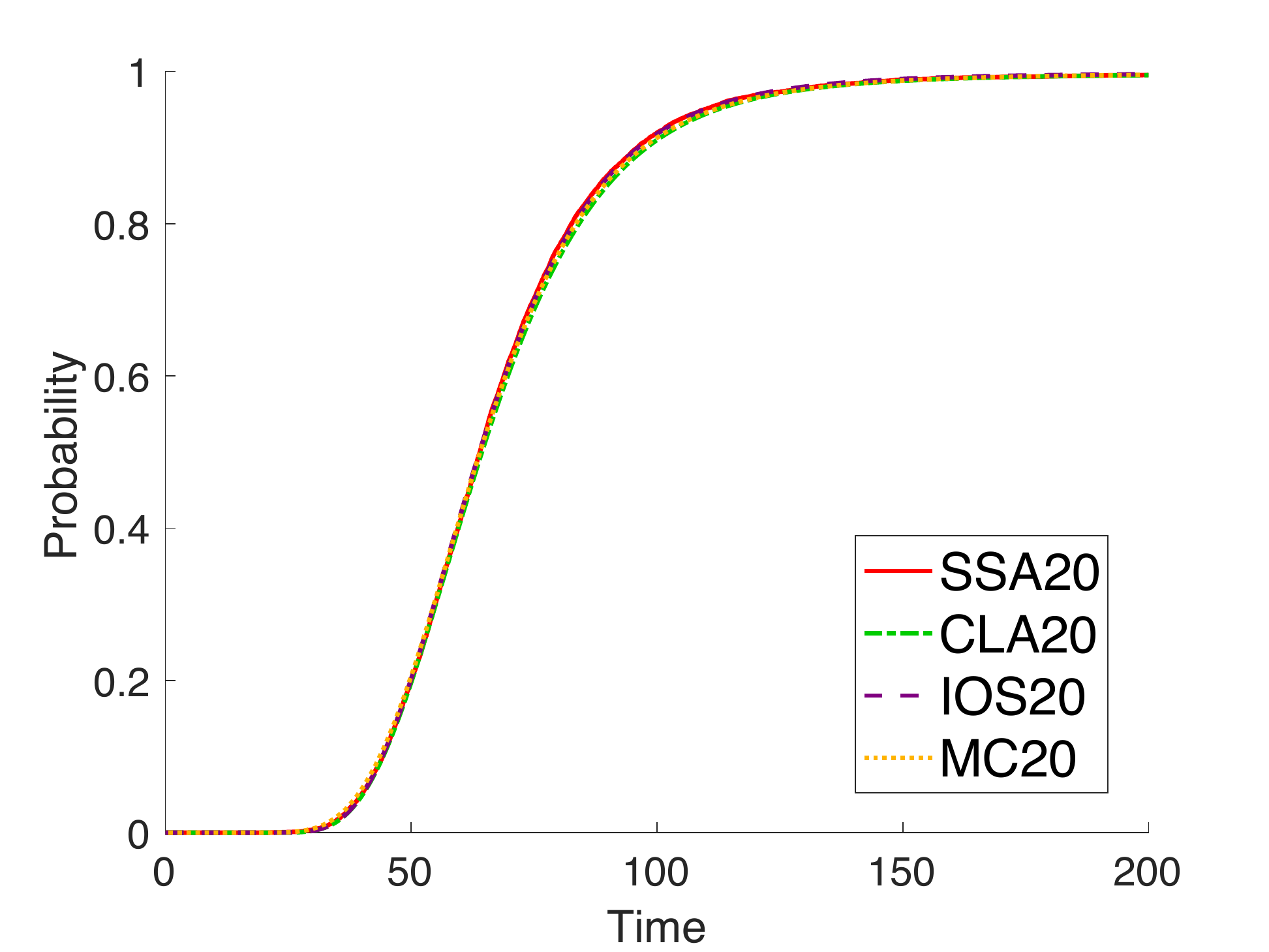}
\end{center}
\caption[Results obtained by System Size Expansion and Moment Closure in the validation of Local-to-Global Properties.] {Comparison of the results obtained by the CLA, the statistical estimate (SSA), the System Size Expansion (IOS), and the Moment Closure (MC) for $N=20$ without (left) and with (right) finite size correction of the threshold, for model parameters $\kappa_{\mathit{inf}}=0.05$, $\kappa_{patch_1}=0.02$, $\kappa_{loss}=0.01$, $\kappa_{ext}=0.05$, $\kappa_{patch_0}=0.001$, $\alpha_1=0.5$ }
\label{fig:highorderMC_IOS}
\end{figure}

\begin{figure}[t]
\begin{center}
\includegraphics[width=.75\textwidth]{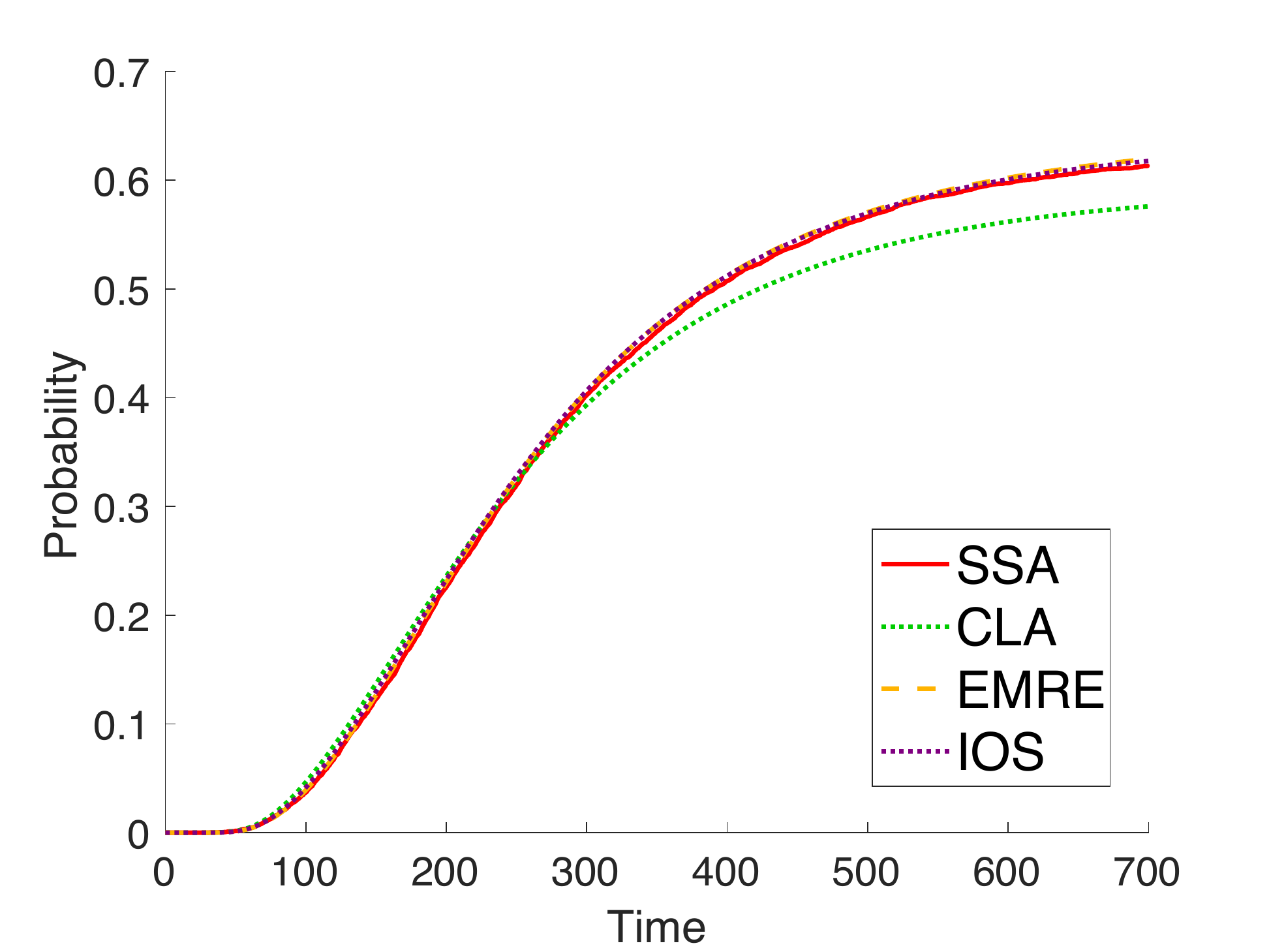}
\end{center}
\caption{Comparison of the results obtained by the CLA, the statistical estimate (SSA), the Effective Mesoscopic Rate Equation (EMRE), and the System Size Expansion (IOS), for $N=20$ and parameters $\kappa_{\mathit{inf}}=0.55$, $\kappa_{patch_1}=0.02$, $\kappa_{loss}=0.01$, $\kappa_{ext}=0.05$, $\kappa_{patch_0}=0.11$, $\alpha_1=0.5$. 
 }
\label{fig:ios}
\end{figure}
\end{example}

As we have seen in the previous example, use of higher-order corrections improves the quality of the estimates. However, this comes with an increased computational cost, which scales as $O(n^k)$, where $n$ is the number of different local states of the agents, and $k$ is the order of the higher order correction. Note that this is still independent from the population size $N$. 
In the future, we plan to stress test the three model checking procedure on more complex property, to understand better the different performances and the quality of the estimations, and their behaviour with respect to the population size $N$. In particular, we want to investigate scenarios where fluid and central limit approximations are  known to perform poorly, like systems exhibiting a multi-stable behaviour. In these cases, we expect higher-order corrections to bring an even more evident gain in accuracy.

\subsection{The Model Checking Algorithm for Collective Properties}

We turn now to discuss how to verify the other collective properties of Definition \ref{def:globalProp}, starting from state properties $\gp{\gprop{\Phi}{a}{b}}{}{\bowtie  p}$.
These are fairly simple to verify, relying on the model checking algorithm for CSL-TA for individual agents discussed in Section \ref{sec:checkCSLTA}.  Essentially, we first run this algorithm and check if the CSL-TA formula $\Phi$ is satisfied, for each state $s\in S$ of an agent class $\class$ at a given initial time $t_0$. Let us call $S(\Phi,t_0) = \{s\in S~|~s,t_0\models \Phi \}$ the set of states satisfying it. Then, checking a path property requires us to compute the probability $P\upsize_{\Phi\in [a,b]}(t_0)$ with which  the variable $X\upsize_{\Phi}(t_0) = \sum_{s\in S(\Phi,t_0)} X\upsize_s(t_0) $
belongs to  $[a,b]$. 
This is difficult to do exactly, but we can rely on the same approximations introduced above for the path probability. More specifically, we consider the basic population model (in this case, there is no need to perform the product construction at the global level, this has already been taken care of while checking the local properties), and compute its moments  either by linear noise approximation or by higher order moment closure techniques. Given the moments of variables $X\upsize_s$, we can easily obtain moments for  $X\upsize_{\Phi}$,\footnote{For the $k$-th non-centred moment, expand the expression $(\sum_{s\in S(\Phi,t_0)} X\upsize_s(t_0) )^k$ and use the values of moments up to order $k$ for the $X_s$ variables.}  from which we can approximate the distribution of $X\upsize_{\Phi}$, either using a Gaussian (for linear noise) or by maximum entropy reconstruction. Finally, we need to integrate numerically this distribution in $[a,b]$ to obtain an approximation $\tilde{P}\upsize_{\Phi\in [a,b]}(t_0)$ of the  probability $P\upsize_{\Phi\in [a,b]}(t_0)$. Verification of $\gp{\gprop{\Phi}{a}{b}}{}{\bowtie  p}$ is concluded by comparing this value with the threshold $p$.

By Theorem \ref{th:convergenceIndividual}, for $N$ large enough the set $S(\Phi,t_0)$ will contain all and only the states satisfying the local specification $\Phi$. Furthermore, if we use the linear noise approximation, we can rely on Theorem \ref{th:central} for the convergence of the distribution of each $X_s$ to its linear noise approximation. The combination of these two results allows us to show that    
 
\begin{theorem}
\label{th:convergence_state}
Under the hypothesis of Theorems \ref{th:central} and \ref{th:convergenceIndividual}, it holds that \linebreak $\lim_{N\rightarrow\infty}\|P\upsize_{\Phi\in [a,b]}(t_0) -  \tilde{P}\upsize_{\Phi\in [a,b]}(t_0)\| = 0 $. \qed
\end{theorem}

This concludes the presentation of the model checking algorithm for collective properties, as Boolean operators are straightforward.

%
%

\section{Conclusion}
\label{sec:conc}

In this paper, we presented a framework for fast and reliable approximate verification of certain classes of properties of population models. In particular, we considered properties of random individuals, referred to as local properties, expressed by the logic CSL-TA (an extension of CSL using Deterministic Timed Automata as temporal modalities), and their lifting to the collective level, computing the probability that the number of agents satisfying the local property meets a given threshold. In order to efficiently compute reachability probabilities, we relied on several stochastic approximations. For individual properties, we exploited the fluid approximation and fast simulation, thus extending fluid model checking \cite{fluidmc} to CSL-TA properties, and we also considered higher-order corrections, exploiting moment closure methods. For the collective properties, we extended the class of properties considered in \cite{qest} to nested CSL-TA specifications, leveraging the central limit approximation and higher order corrections combined with maximum entropy distribution reconstruction routines.  For both classes of properties, we provided  theoretical results guaranteeing convergence in the limit of infinite populations, and experimental evidence of the effectiveness of the method.
From a practical point of view, the approach we presented is computationally efficient, outperforming even statistical model checking, while being accurate already for populations of moderate size. Furthermore, its complexity depends only on the number of local states and transitions of a model, and not on the population level, and convergence results guarantee that the error decreases as the population increases. Hence, this approach is very effective for medium and large populations, on the order of hundreds of individuals or more, precisely when exact and statistical methods start to suffer from a prohibitive large computational cost.  

This work can be extended in few directions. First, by providing a tool taking care of automatising the steps required to check a property. Secondly, by considering a more general class of local properties, i.e. removing the restriction that clocks cannot be reset. At the individual level, we can rely on the results of  \cite{formats15}, building on top of the fluid approximation of population models with deterministic time delays \cite{qest12,hayden}. In  \cite{formats15}, clock resets  introduce deterministic delays in model of an individual agent synchronised with a property, reflecting into the fluid equations and the Kolmogorov equations for individuals becoming Delay Differential Equations. 
 The challenge with these Delay Differential Equations is their stiffness, which calls for effective numerical solving routines to make them usable in practice. Lifting to the collective level requires a central limit approximation, which can be crafted building on the results of \cite{hayden}. Moment closure  for this class of models, instead, is still an open research problem. 
 Finally, probably the most challenging and rewarding direction to investigate is that of providing tight error bounds that could be used to assess when the approximation is good and when it may be questionable. This is  challenging, as known error bounds even for fluid approximation tend to be over-conservative, being based on worst-case inequalities like Gronwall's one \cite{kurtz}.  A possible direction is to generalize and exploit the approach of \cite{bortolussi2013}.


\section*{Acknowledgements}
This research has been partially funded by the EU-FET project QUANTICOL (nr. 600708).


\bibliographystyle{elsarticle-num}
\bibliography{biblio}


\clearpage
\appendix

\section{Proofs}
\label{app:proofs}

In this appendix, we provide the proofs of the main results of the paper.

\bigskip

\noindent\textbf{Proposition~\ref{prop:individualRates}.}
The rate of transition $(s, q) \xrightarrow{\loclab{s}} (s', q')$  of an individual agent due to global transition $\tau$, given that the population model is in state $\X\upsize(t) = \x$, is
\[ g\upsize_{\tau}((s,q),(s',q')) = \frac{m_\tau}{x_s}  f\upsize_{\tau}(\x).\]

\proof To obtain the expression for  $g\upsize_{\tau}((s,q),(s',q'))$, we need to compute which is the probability that the tagged agent is one of the randomly chosen agents in state $s$ that are updated by the transition $\tau$, and multiply the global rate of a $\tau$ transition by it. As there are $x_s$ agents in state $s$, and $m_\tau$ are involved in the transition, this probability is readily computed as the fraction of subsets of $m_\tau$ elements of a set of  $X_s$ elements that contain a fixed element.  This number is
\[ \frac{\binom{x_s-1}{m_\tau-1}}{\binom{x_s}{m_\tau}} = \frac{m_\tau}{x_s}. \]

\qed

\bigskip

\noindent\textbf{Proposition~\ref{prop:global_rates}.}
With the definitions of Section \ref{sec:collectivesynch}, it holds that
$\sum_{\vec{q}\in Q^k} f_{\vec{q}}\upsize (\pstvec) = f\upsize (\pstvec)$, i.e.
\[ \sum_{\vec{q}\in Q^k}  \frac{\prod_{(s,q)\in LHS(\syncset{\tau,\vec{q}})} \frac{X_{s,q}!}{(X_{s,q}-\kappa_{s}^{q} )!}}{\prod_{s\in LHS(\syncset{\tau})} \frac{X_{s}!}{(X_{s}-\kappa_{s} )!}}  = 1\]

\proof  We start by providing a more detailed derivation of the formula for the rate $f_{\vec{q}}\upsize (\pstvec)$. As stated in the main text, we fix an ordering of the elements in $\syncset{\tau}$, and count how many ordered tuples we can construct in the aggregated state $\widetilde{\pstvec} = (X_1, \ldots, X_n)$, where $X_s = \sum_{r = 1}^{m} X_{s,r}$. Hence, the element $j$ of the so build tuple is an agent in state $s_j$, where $s_j$ is the left hand side of the $j$-th update rule in $\syncset{\tau}$. Now, if state $s$ appears $\kappa_{s}$ times in the lhs of a rule in $\syncset{\tau}$, then each time it appears we pick an agent of type $X_s$. The first time there are $X_s$ possible choices, the second time $X_s-1$ and so on. It follows that the contribution of agents in state $s$ to the number of $\syncset{\tau}$-tuples is $X_s(X_s-1)\cdots (X_s-\kappa_s +1) = \prod_{h< \kappa_s}(X_s-h) = \frac{X_{s}!}{(X_{s}-\kappa_{s} )!}$.  Here we are implicitly assuming $X_s \geq \kappa_s$, and set to zero such product otherwise. To count the number of tuples, we now have to multiply these expressions for each state $s$ appearing in the lhs of a rule of $\syncset{\tau}$. What we get is the denominator  $\prod_{s\in LHS(\syncset{\tau})} \frac{X_{s}!}{(X_{s}-\kappa_{s} )!}$. The numerator is computed similarly, just considering the rules in $\syncset{\tau,\vec{q}}$, for a fixed $\vec{q}$. \\
Now, to prove the formula, consider agents in the product model, with states $(s,q)$, and build a $\syncset{\tau}$-tuples with them, ignoring the part of the state coming from the property, i.e. $q$. Each agent in state $s$ in such a tuple will nonetheless have also a property state $q$ associated with it. If we enumerate such property states $q$ for all the agents in the tuple, we get a vector $\vec{q}$. In this way, we can assign each such a tuple to one vector $\vec{q}$, hence partitioning the set of $\syncset{\tau}$-tuples  built ignoring the property state into disjoint subsets, one for each $\vec{q}$. It is clear that if we count how many tuples of type $\vec{q}$ there are, and we add this number for each $\vec{q}$, we are counting the cardinality of all the $\syncset{\tau}$-tuples. Due to the discussion above, the number of tuples of type $\vec{q}$ is $\prod_{(s,q)\in LHS(\syncset{\tau,\vec{q}})} \frac{X_{s,q}!}{(X_{s,q}-\kappa_{s}^{q} )!}$, and the number of $\syncset{\tau}$-tuples is $\prod_{s\in LHS(\syncset{\tau})} \frac{X_{s}!}{(X_{s}-\kappa_{s} )!}$. Hence
\[ \sum_{\vec{q}\in Q^k} \prod_{(s,q)\in LHS(\syncset{\tau,\vec{q}})} \frac{X_{s,q}!}{(X_{s,q}-\kappa_{s}^{q} )!} = \prod_{s\in LHS(\syncset{\tau})} \frac{X_{s}!}{(X_{s}-\kappa_{s} )!}. \] \qed

\bigskip

\noindent\textbf{Theorem~\ref{th:convergence}.}
Under the hypothesis of Theorem \ref{th:central}, it holds that \linebreak $\lim_{N\rightarrow\infty}\|{P}\upsize_\gdta(T) -  \tilde{P}\upsize_\gdta(T)\| = 0 $. 

\proof 
Recall that by Theorem \ref{th:central}, the sequence of random processes $\linear\upsize(t)\ :=\ \size^{\frac{1}{2}}\left(\pnstvec\upsize(t) - \fluid(t)\right)$ converges to the Gaussian random process $\linear(t)$ obtained by the central limit approximation. Assume for the moment that the population model associated with the property $\gdta$ is composed by a single model. It is easy to verify that the conditions to apply Theorem \ref{th:central}  are satisfied (all rate functions of the modified population models are Lipschitz continuous). In particular, the initial conditions  for $\linear\upsize(t)$ and $\linear(t)$ converge by definition.
 As we are interested in the value of those processes at a fixed time $T>0$, let $\linear\upsize = \linear\upsize(T)$ and $\linear = \linear(T)$. Theorem \ref{th:central} implies that $\linear\upsize \Rightarrow \linear$ (weak convergence).

First of all, we transform the interval $[a,b]$ into a $\size$-dependent interval $[a\upsize,b\upsize]$, so that we can evaluate ${P}\upsize_\gdta(T)$ as $\bbP\{Z_{Final}\upsize\in [a\upsize,b\upsize]\}$ and  $\tilde{P}\upsize_\gdta(T)$ as $\bbP\{Z_{Final} \in [a\upsize,b\upsize]\}$, where $Z_{Final}\upsize$ and $Z_{Final}$ are the marginal distributions of $\linear\upsize$ and $\linear$ on the coordinate corresponding to $X_{Final}$. By the definition of $\linear\upsize$, it easily follows that $a\upsize = \size^{\frac{1}{2}}\left(a - \fluid_{Final}(T)\right)$ and $b\upsize = \size^{\frac{1}{2}}\left(b - \fluid_{Final}(T)\right)$.

Ideally, to prove the convergence of the probability values, we would like to invoke the Portmanteau theorem\footnote{See P. Billingsley. Convergence of Probability Measures, 2nd edition. Wiley, 1999.}, using the weak convergence of $\linear\upsize$ to $\linear$. However, this does not work here, as the sets for which we have to evaluate the probability depend on $\size$. Hence, we need a slightly trickier argument.
By the triangular inequality, we have
\[  
\begin{split}
\|   \bbP\{Z_{Final}\upsize\in [a\upsize,b\upsize]\} & - \bbP\{Z_{Final}\in [a\upsize,b\upsize]\} \| \leq \\
& \underbrace{\|   \bbP\{Z_{Final}\upsize\in [a\upsize,b\upsize]\} - \bbP\{Z_{Final}\upsize\in [a^\infty,b^\infty]\} \|}_{(a)} +\\ 
& \underbrace{\|   \bbP\{Z_{Final}\in [a^\infty,b^\infty]\} - \bbP\{Z_{Final}\in [a\upsize,b\upsize]\} \|}_{(b)}
\end{split}
\]
where   $[a^\infty,b^\infty]$ is the the limit set to which $[a\upsize,b\upsize]$ converges as $\size$ goes to infinity. Clearly, $a^\infty = \lim_{N\rightarrow\infty} a\upsize$, and similarly for $b^\infty$. We have four cases, depending on the relative value of $a$ and $b$ with respect to $\fluid_{Final}(T)$:
\begin{enumerate}
\item if $a,b > \fluid_{Final}(T)$ or $a,b < \fluid_{Final}(T)$, then $[a^\infty,b^\infty] = \emptyset$. In fact, in the first case, both $a^\infty = +\infty$ and $b^\infty = +\infty$;
\item if $a< \fluid_{Final}(T)$ and $b > \fluid_{Final}(T)$, then $[a^\infty,b^\infty] = [-\infty,+\infty] = \bbR$;
\item if $a =  \fluid_{Final}(T)$ and $b > \fluid_{Final}(T)$, then $[a^\infty,b^\infty] = [0,+\infty]$; 
\item if $a< \fluid_{Final}(T)$ and $b = \fluid_{Final}(T)$, then $[a^\infty,b^\infty] = [-\infty,0]$;
\end{enumerate}

The term (b) in the inequality above goes to zero, due to convergence of $[a\upsize,b\upsize]$ to $[a^\infty,b^\infty]$. To deal with term (a), instead, we can exploit the fact that, as $Z_{Final}\upsize\Rightarrow Z_{Final}$ and $\bbR$ is a Polish space, by the Prohorov theorem, $Z_{Final}\upsize$ is uniformly tight, hence for each $\eps>0$ there is $k_\eps>0$ such that, for all $\size$, $\bbP\{Z_{Final}\upsize \in [-k_\eps,k_\eps]\}>1-\eps$. We deal with the four cases above separately:
\begin{enumerate}
\item Fix $\eps>0$ and let $N_0$ be such that, for $N\geq N_0$, $[a\upsize,b\upsize]\cap [-k_\eps,k_\eps] = \emptyset$. It follows that $\bbP\{Z_{Final}\in [a\upsize,b\upsize]\} < \eps$. As $\bbP\{Z_{Final}\upsize\in [a^\infty,b^\infty]\} = 0$, the term (a) is less than $\eps$, which implies that (a) goes to zero for $N$ going to infinity. 
\item Fix $\eps>0$ and let $N_0$ be such that, for $N\geq N_0$, $[a\upsize,b\upsize]\cap [-k_\eps,k_\eps] = [-k_\eps,k_\eps]$. As $\bbP\{Z_{Final}\upsize\in [a^\infty,b^\infty]\} = 1$, it follows that (a) is smaller than $\eps$, hence it has limit 0.
\item Fix $\eps>0$ and let $N_0$ be such that, for $N\geq N_0$, $[a\upsize,b\upsize]\cap [-k_\eps,k_\eps] = [0,k_\eps]$. By the monotonicity of the probability distributions, term (a) is smaller than $\bbP\{Z_{Final} > k_\eps\}$, which is itself smaller than $\eps$. Also in this case, it follows that (a) has limit 0. 
\item This case is symmetric with respect to case 3.
\end{enumerate}

Putting all together, we have shown that $\|   \bbP\{Z_{Final}\upsize\in [a\upsize,b\upsize]\}  - \bbP\{Z_{Final}\in [a\upsize,b\upsize]\} \|$ goes to zero as $N$ goes to infinity, as desired. 

In order to deal with the cases in which  the population model associated with the property $\gdta$ is a sequence of $k>1$ models, we can rely on the fact that the time constants defining intervals $I_j$ are fixed, hence Theorem \ref{th:central} holds inductively for each model of the sequence. In fact, the initial conditions for model $\popn_{I_{j}}$ are given by the  final state of model $\popn_{I_{j-1}}$, which converge by inductive hypothesis. Therefore, we just need to apply the argument discussed above to the final model of the sequence. \qed

\end{document}